\documentclass[times,5p,twocolumns,final,numbers]{elsarticle}

\usepackage{graphicx}
\usepackage{subfig}
\usepackage{tabularx}
\usepackage{hyperref}
\usepackage{color}
\usepackage{xcolor}

\usepackage{booktabs}
    \newcolumntype{L}{>{\raggedright\arraybackslash}X}
    
\usepackage{forest}
\usepackage{amsmath,bm,times}

\journal{Journal of \LaTeX\ Templates}

\begin{document}

\begin{frontmatter}

\title{Study on Transfer Learning Capabilities for Pneumonia Classification\\ in Chest-X-Rays Images}

\author[1]{Danilo Avola}
\author[1]{Andrea Bacciu}
\author[1]{Luigi Cinque}
\author[1]{Alessio Fagioli\corref{cor1}}
\cortext[cor1]{Corresponding author: 
	Tel.: +39 06 8576 8425;}
\ead{fagioli@di.uniroma1.it}
\author[1]{Marco Raoul Marini}
\author[1]{Riccardo Taiello}

\address[1]{Department of Computer Science, Sapienza University, Via Salaria 113, 00185, Rome, Italy}

\begin{abstract}

\textit{Background:} 
over the last year, the severe acute respiratory syndrome coronavirus-2 (SARS-CoV-2) and its variants have highlighted the importance of screening tools with high diagnostic accuracy for new illnesses such as COVID-19. In that regard, deep learning approaches have proven as effective solutions for pneumonia classification, especially when considering chest-x-rays images. 
However, this lung infection can also be caused by other viral, bacterial or fungi pathogens.
Consequently, efforts are being poured toward distinguishing the infection source to help clinicians to diagnose the correct disease origin. 
Following this tendency, this study further explores the effectiveness of established neural network architectures on the pneumonia classification task through the transfer learning paradigm.

\textit{Methodology:} 
to present a comprehensive comparison, 12 well-known ImageNet pre-trained models were fine-tuned and used to discriminate among chest-x-rays of healthy people, and those showing pneumonia symptoms derived from either a viral (i.e., generic or SARS-CoV-2) or bacterial source.
Furthermore, since a common public collection distinguishing between such categories is currently not available, two distinct datasets of chest-x-rays images, describing the aforementioned sources, were combined and employed to evaluate the various architectures.

\textit{Results:} 
the experiments were performed using a total of 6330 images split between train, validation, and test sets. For all models, standard classification metrics were computed (e.g., precision, f1-score), and most architectures obtained significant performances, reaching, among the others, up to 84.46\% average f1-score when discriminating the four identified classes. Moreover, execution times, areas under the receiver operating characteristic (AUROC), confusion matrices, activation maps computed via the Grad-CAM algorithm, and additional experiments to assess the robustness of each model using only 50\%, 20\%, and 10\% of the training set were also reported to present an informed discussion on the networks classifications.

\textit{Conclusion:} this paper examines the effectiveness of well-known architectures on a joint collection of chest-x-rays presenting pneumonia cases derived from either viral or bacterial sources, with particular attention to SARS-CoV-2 contagions for viral pathogens; demonstrating that existing architectures can effectively diagnose pneumonia sources and suggesting that the transfer learning paradigm could be a crucial asset in diagnosing future unknown illnesses. 

\end{abstract}


\end{frontmatter}

\begin{figure*}[t]
    \centering
	\subfloat[]{\includegraphics[scale=0.595]{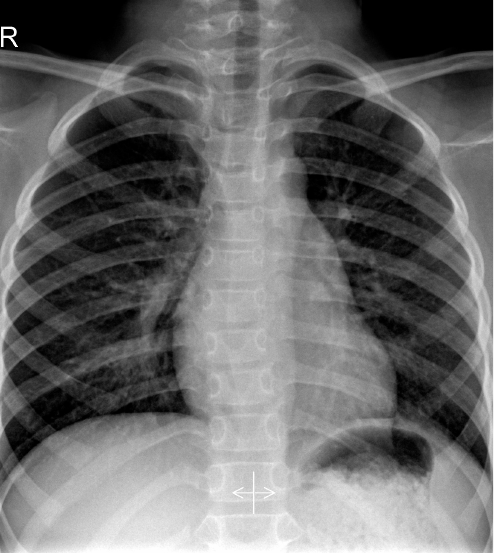}}
	\hfil
	\subfloat[]{\includegraphics[scale=0.595]{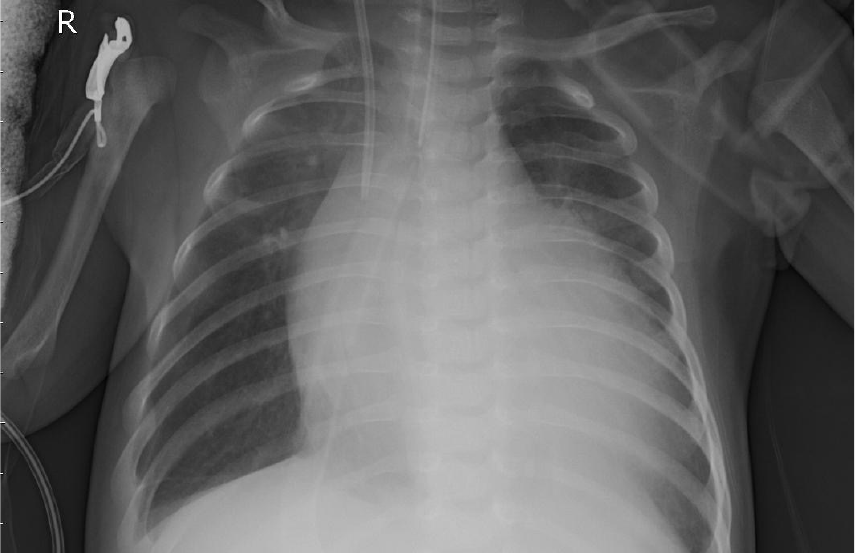}}
	\hfil
	\subfloat[]{\includegraphics[scale=0.595]{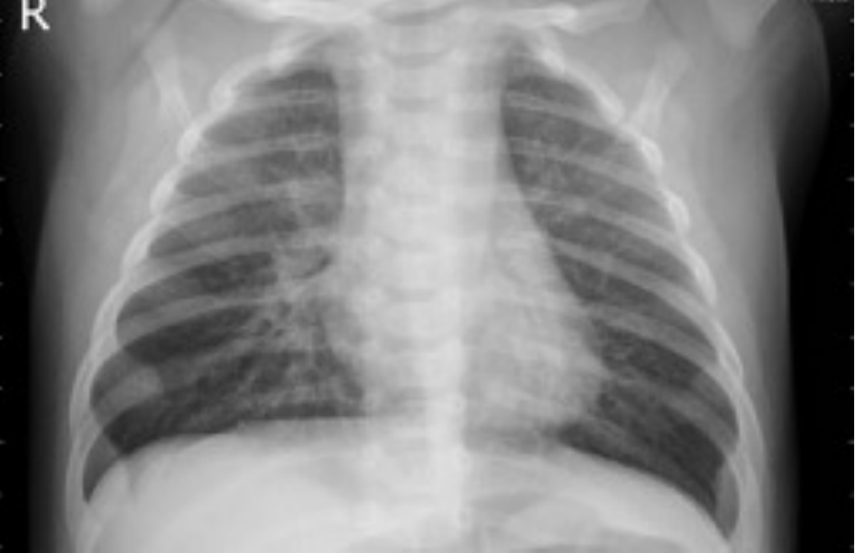}}
	\hfil
	\subfloat[]{\includegraphics[scale=0.595]{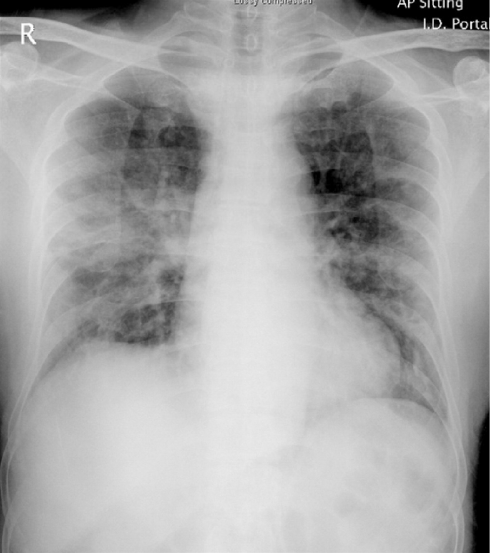}}
    \caption{Chest-x-ray samples showing a healthy patient (a), pneumonia from a bacteria infection (b), pneumonia from a virus infection (c), and pneumonia from a COVID-19 case (d).}
    \label{fig:pneumonia_x_rays}
\end{figure*}

\section{Introduction}
Pneumonia is an acute respiratory infection caused by viral, bacterial or fungal pathogens. This infection can affect either a single or both lungs and can cause mild to life-threatening effects in people of all ages. Notably, there are increased risks for those with preexisting health conditions or in adults with more than 65 years, and is the largest worldwide infectious cause of death for children; accounting for 15\% of children deaths in 2017, according to World Health Organization (WHO) statistics. Among the possible sources, viruses and bacteria are the most common pneumonia infection causes \cite{smith2014trends} and, depending on the pathogen, the illness shows a different behavior. In particular, for viral agents, the infection is generally diffused across all lungs \cite{shah2017does}, while for bacterial ones it usually concentrates on given areas \cite{reynolds2010pneumonia}. Therefore, due to intrinsic differences between viruses and bacteria, as well as their potential treatment, a key aspect is to correctly diagnose the pneumonia source. The latter has become even more crucial since the discovery of a novel coronavirus in 2019 \cite{huang2020clinical}, that has resulted in the SARS-CoV-2 virus and related COVID-19 disease pandemic outbreak. 
Indeed, while many works are addressing this diagnosis task \cite{bhattacharya2021deep}, there is still a need for increasingly more general research that might be used to oppose the possibly already endemic SARS-CoV-2 \cite{phillips2021coronavirus}, its future variants \cite{fontanet2021sars}, or completely new diseases that might result in other pandemics \cite{michie2021sustained}.

To address the pneumonia classification task and help clinicians with their diagnoses, an effective solution is to analyze chest-x-rays images. Indeed, as can be observed in Fig.~\ref{fig:pneumonia_x_rays}, the infection spread can be clearly visible through the x-ray technology, especially for severe conditions such as the one associated to the SARS-CoV-2 virus. To fully exploit information found in such images, deep learning approaches are being explored since they can already obtain significant performances in several heterogeneous medical fields \cite{william2018review,karthik2020neuroimaging,marti2020survey,avola2021ultrasound}. In detail, many solutions focus on the binary classification between COVID-19 patients and healthy people \cite{minaee2020deep,brunese2020explainable,shaban2021detecting,tabik2020covidgr,zhang2020viral,shamsi2021uncertainty}. For instance, the authors of \cite{shaban2021detecting} design a custom deep neural network (DNN) model with a hybrid diagnose strategy (HDS) and fuzzy inference engine to identify COVID-19 patients, while in \cite{shamsi2021uncertainty} transfer learning is applied to 4 pre-trained networks to distinguish between SARS-CoV-2 infected and healthy people. Although a binary classification allows a given model to completely focus on the COVID-19 disease, several works try to build more general models by also differentiating pneumonia patients with other sources (i.e., bacterial or virus) through the use of deep learning methods \cite{demir2021deepcoronet,ozturk2020automated,gupta2021instacovnet,tang2021edl,saleh2021coronavirus,nour2020novel}. In \cite{gupta2021instacovnet}, for example, the authors devise a stacking ensemble method to leverage several ImageNet pre-trained architectures and discern between healthy, COVID-19 and other pneumonia-affected patients; while a custom architecture based on long-short term memory (LSTM) units is trained from scratch in \cite{demir2021deepcoronet} to achieve a similar purpose. Finally, different works propose yet another abstraction step by recognizing 4 different classes, i.e., healthy, COVID-19, as well as pneumonia patients infected via bacterial or other viral pathogens \cite{khan2020coronet,oh2020deep,li2021multiscale,ahmad2021deep,mahmud2020covxnet,karthik2021learning}. In particular, the authors of \cite{khan2020coronet} introduce a convolutional neural network (CNN), called CoroNet, specifically designed and trained to discriminate between the aforementioned categories. In \cite{ahmad2021deep}, instead, 5 ImageNet pre-trained architectures are combined via the ensemble technique to increase their system classification performance. 

An important aspect for methods addressing the pneumonia classification in COVID-19 affected patients and, more in general, of new illnesses is the lack of datasets, especially when diseases are first discovered. In this context, approaches based on few-shot learning (FSL), where models are trained on an extremely small number of annotated samples, can be developed and prove useful as early diagnosis tools \cite{chen2021momentum,abdel2021fss,voulodimos2021few}. In \cite{chen2021momentum}, for instance, an encoder is trained over a small number of samples to extract image embeddings via contrastive learning. This low-dimensional representation is then fed to a prototypical network that can diagnose the COVID-19 infection. The authors of \cite{abdel2021fss}, instead, exploit FSL to train in a self-supervised way an architecture organized as an encoder-decoder and perform segmentation of CT scans to detect a SARS-CoV-2 contagion more easily. Although FSL can considerably help diagnose newfound illnesses, a relevant problem that can arise when using few samples is the privacy of the corresponding patients, which should always be guaranteed. To this end, several approaches exploit the federated learning (FL) technique when implementing their solution to diagnose COVID-19 \cite{dayan2021federated,kumar2021blockchain,yan2021experiments}. In particular, FL enables devices to collaboratively learn a shared global model without sharing their local training data with a single server, therefore ensuring data privacy through a decentralized approach. 
The authors of \cite{kumar2021blockchain}, for example, address the segmentation and classification of COVID-19 while preserving user privacy by developing a capsule network that is trained through the blockchain FL. Differently, the scheme in \cite{yan2021experiments} presents a custom strategy where a shared model can maintain data without physically exchanging them and train several known networks on the COVID-19 diagnosis using the developed methodology.

\begin{figure}
    \centering
    \includegraphics[width=\columnwidth]{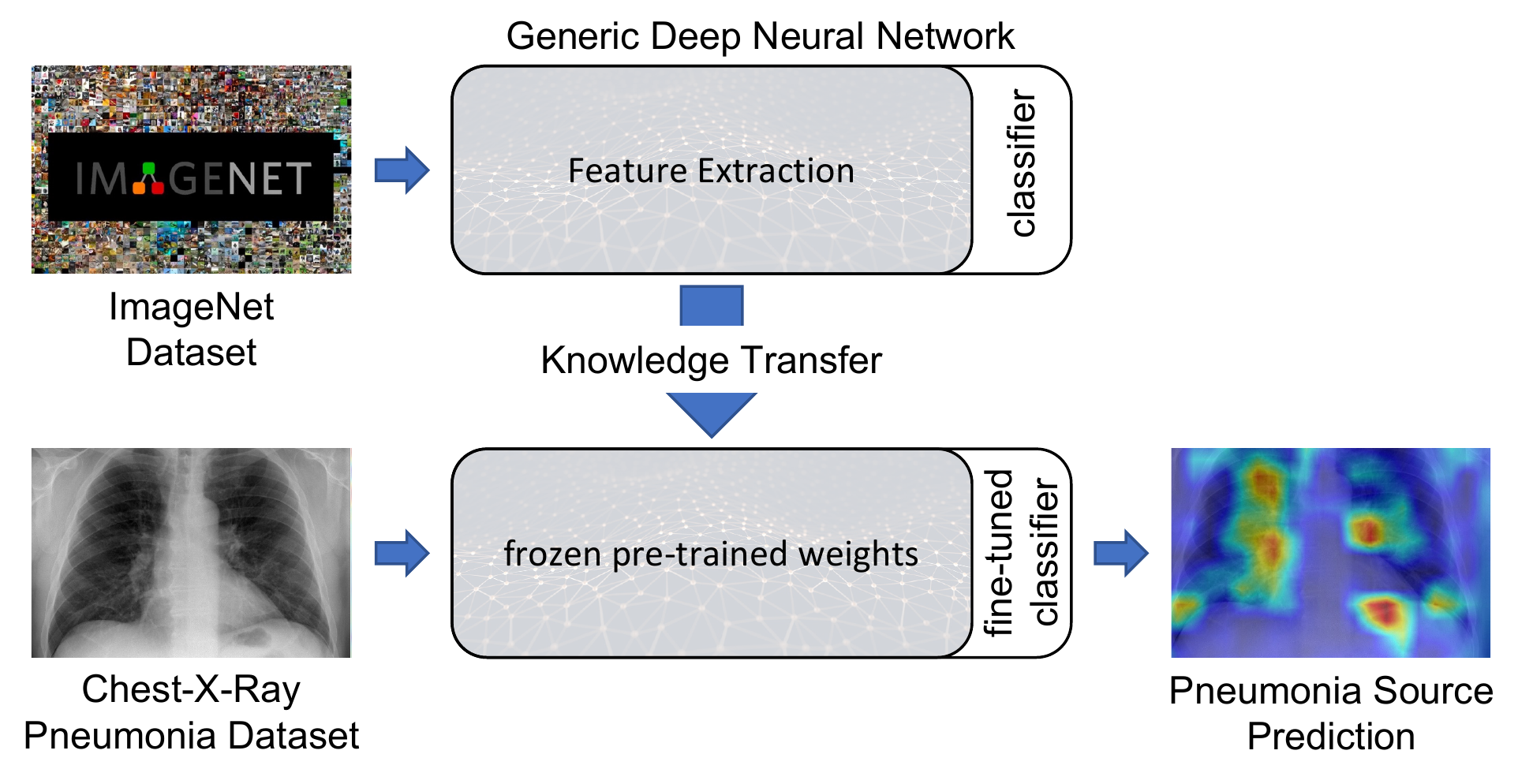}
    \caption{Transfer learning paradigm scheme for pneumonia classification.}
    \label{fig:transfer_learning}
\end{figure}
While many literature works focusing on the SARS-CoV-2 infection introduce specific architectures that handle its  classification, or design models that can keep learning following the continual learning paradigm  \cite{wang2021does,lee2020clinical,vokinger2021continual}, others simply focus on transfer learning capabilities by fine-tuning known architectures on pneumonia datasets \cite{marques2020automated,minaee2020deep,katsamenis2020transfer,apostolopoulos2020covid,aslan2021cnn,pathak2020deep,horry2020covid}. Following the latter methodology, we performed a transfer learning study by fine-tuning 12 models on an extended collection derived by combining two public datasets presented, respectively, in \cite{kermany2018identifying} and \cite{minaee2020deep}. In particular, we froze all Imagenet pre-trained models up to their respective classification component, as summarized in Fig.~\ref{fig:transfer_learning}, which was instead modified to handle the classification task of the 4 available categories, i.e., healthy and pneumonia derived from bacterial, viral and SARS-CoV-2 pathogens.
This allows to train even complex models in less time and enables the presented architectures to be ready to handle a complex task such as medical image classification. Furthermore, we also present qualitative experiments to explain the networks choices by directly observing what they focus on inside the chest-x-ray images via the Grad-CAM algorithm \cite{selvaraju2017grad}. Presenting such a comprehensive transfer learning overview helps to choose better backbone architectures to build custom models upon, and offers a strong methodology for possible future diseases which might require to be diagnosed as early as possible, and for which it might be difficult to design good performing neural networks right away.

The rest of this paper is structured as follows. A brief introduction on the rationale behind each pre-trained model used in this work is provided in Section~\ref{sec:method}. The dataset, quantitative and qualitative experimental results, as well as a discussion on the obtained performances, are presented in Section~\ref{sec:experimental_results}. Finally, work conclusions and future plans are drawn in Section~\ref{sec:conclusion}.

\section{Materials and Methods}\label{sec:method}

The intuition behind transfer learning is to use a model pre-trained on a large and heterogeneous dataset, so that its generic world representation can be exploited on a new and different task \cite{pan2009survey}. In particular, when analyzing images, this learning approach takes advantage of feature maps learned by a network, without requiring to train the model from scratch and, therefore, avoiding a time-consuming and resource-intensive procedure. To transfer previous knowledge to the new task, there are two possible approaches. In the first one, a pre-trained model is treated as a feature extractor by freezing its internal weights. A classifier is then trained on top of this frozen architecture to achieve the knowledge transferal, and retain as much information as possible from the previous field. Notice that the classifier is usually implemented either via one or more dense layers, with the last one using the softmax function to obtain the probability distribution over the available classes; as a completely distinct deep learning architecture; or through machine learning algorithms such as the support vector machine (SVM) or random forest (RF). In the second approach, instead, the whole architecture, or a subset thereof, is fine-tuned on the new task. 
In this case, the pre-trained model weights are used as a baseline instead of starting from a random initialization. This generally allows a network to reach training convergence faster, and to specialize its entire structure to the new field. While both methodologies can be effective, in this work, we explore the former approach and assess the capabilities of various networks used as feature extractors; and train the original classifier component of each model, which is implemented as one or more fully connected layers, on the collected dataset. This training strategy allows to emulate the scarcity of data, typical when new illnesses are first discovered, but still enables to correctly train otherwise data hungry networks \cite{cheplygina2019not}. In more detail, we performed experiments on a combined collection derived from datasets presented in \cite{kermany2018identifying} and \cite{minaee2020deep} to evaluate 12 well-known ImageNet pre-trained models briefly discussed later in this section, i.e., AlexNet \cite{krizhevsky2012imagenet}, DenseNet \cite{huang2017densely}, GoogleNet \cite{szegedy2015going}, MnasNet \cite{tan2019mnasnet}, MobileNetv2 \cite{sandler2018mobilenetv2}, MobileNet v3 (Large) \cite{howard2019searching}, ResNet50 \cite{he2016deep}, ResNeXt \cite{xie2017aggregated}, ShuffleNet \cite{zhang2018shufflenet}, SqueezeNet \cite{iandola2016squeezenet}, VGG16 \cite{simonyan2014very} and Wide ResNet50 \cite{zagoruyko2016wide}. 
Moreover, each model $\mathcal{M}$ was modified and adapted to the new task by changing the output layer $\mathcal{M}_o$ nodes number, by exploiting the following equation:
\begin{equation}
    \mathcal{M}_{logits} = \textbf{W}_{o} * Dropout(ReLU(\textbf{W}_h \mathbf{x})),
\end{equation}
where $\mathcal{M}_{logits}$ and $\textbf{W}_o$ represent the model output logits and $\mathcal{M}_o$ weights, respectively;  $Dropout(\cdot)$ is a regularization technique employed to avoid overfitting; $ReLU(\cdot)$ corresponds to the chosen activation function applied to the incoming inputs; $\textbf{W}_h$ indicates weights between $\mathcal{M}_{o}$ and its preceding hidden layer  $\mathcal{M}_h$; while $\mathbf{x}$ denotes the $\mathcal{M}_h$ output. In particular, given the model-dependent input size $n=|\mathbf{x}|$, $\textbf{W}_h$ and $\textbf{W}_{o}$ will have a dimension of $\mathbb{R}^{n \times 256}$ and $\mathbb{R}^{256 \times 4}$, respectively. Finally, to train a model $\mathcal{M}$, we employed the categorical cross-entropy loss $\mathcal{L}$ defined as:
\begin{equation}
    \mathcal{L} = - \frac{1}{N}\sum_{i=0}^{N-1} y_i \log \hat{y}_i,
\end{equation}
where $N$ corresponds to the number of classes, i.e., normal, bacteria, virus, and COVID-19; while $\hat{y}_i$ and $y_i$ indicate, respectively, the predicted output and ground truth labels. 

\begin{table}[t]
	\centering
	\resizebox{\columnwidth}{!}{
	\begin{tabular}{lccccccc}
		\toprule
		& Bacteria & Normal & Virus & \multicolumn{2}{c}{COVID-19} & Total  \\
		\cmidrule(lr){2-4}
		\cmidrule(lr){5-6}
		& \multicolumn{3}{c}{\cite{kermany2018identifying}} & \cite{minaee2020deep} & \cite{cohen2020covid,cohen2020covidProspective}$^1$ & \\
		\midrule
		Train & 2249 & 1060 & 1056 & 84  & -   & 4449 \\
		Val   & 289  & 289  & 289  & -   & 289 & 1156 \\
		Test  & 242  & 234  & 148  & 101 & -   & 725  \\
		\bottomrule
	\end{tabular}%
	}
	\footnotesize{$^1$https://github.com/ieee8023/covid-chestxray-dataset}
	\caption{Final merged dataset distribution.}
	\label{tab:dataset}
\end{table}

\subsection{AlexNet}
The AlexNet \cite{krizhevsky2012imagenet} architecture had a significant impact in the computer vision field over the last decade. Its success can be related to the ImageNet LSVRC-2012 \cite{russakovsky2015imagenet} competition, which it won by a large margin compared to the second classified, i.e., with an accuracy increase of 10.9\%. The architecture comprises $5$ convolutional layers using kernels with size 11$\times$11, 5$\times$5, and 3$\times$3 for the last three layers. Moreover, max pooling, dropout, and ReLU are applied after the first, second, and fifth convolution to extract a vector of relevant features with dimension 9216. Finally, the latter are classified through $3$ fully connected layers with a shape of 4096, 4096, and 1000, respectively.

\subsection{DenseNet} 

A more recent model is the DenseNet \cite{huang2017densely}, developed to mitigate the vanishing/exploding gradient problem which is a typical issue in deep neural networks with many layers. To this end, the authors organized their network into dense blocks interleaved by transition layers that can act as a bottleneck to reduce the architecture parameters number as well as the input size. Moreover, a dense block can contain several convolutions (e.g., 6, 12, 24), and each of these operations receives as input all of the feature maps produced by its preceding layers. This allows to forward more information inside a given block, and results in an overall more compact configuration, via this dense blocks subdivision, as well as in improved performances. For this model, the classifier is composed of a single dense layer taking as input a feature vector of size 1024.

\subsection{GoogLeNet}
Another key architecture is GoogLeNet \cite{szegedy2015going}, organized into 27 layers containing convolutions as well as $5$ max pooling operations and $9$ inception modules. The latter are designed to allow a more efficient computation through a dimensionality reduction achieved via 1$\times$1 convolutions. What is more, on top of these convolutions, GoogLeNet also retains crucial spatial information by analyzing the input at different scales by using kernel sizes of 1$\times$1, 3$\times$3, and 5$\times$5 inside its inception modules. Finally, intermediate classifications are also performed to reinforce the final network output and improve the network gradient throughout its training. To this end, while the classifier of this model receives as input a feature vector of dimension 1024, the intermediate classifiers are fed with vectors of size 2048.

\subsection{MnasNet}

A network specifically optimized to obtain significant performances on mobile devices through the Mobile Neural Architecture Search (MNAS) approach, is the MnasNet \cite{tan2019mnasnet}. In particular, the architecture latency, i.e., the time required to perform inference on a given image, is incorporated directly into the main objective to be minimized at training time; enabling the model search to find a good trade-off between accuracy and latency. Moreover, the MnasNet employs $9$ layers comprising convolutions as well as depthwise separable convolutions and mobile inverted bottlenecks, which exploit inverted residual connections, to extract a feature vector with 1280 elements that is classified using a single dense layer. Notice that both techniques were defined by the MobileNet v2, presented in the following section.

\begin{figure}[t]
    \centering
	\includegraphics[width=.8\columnwidth]{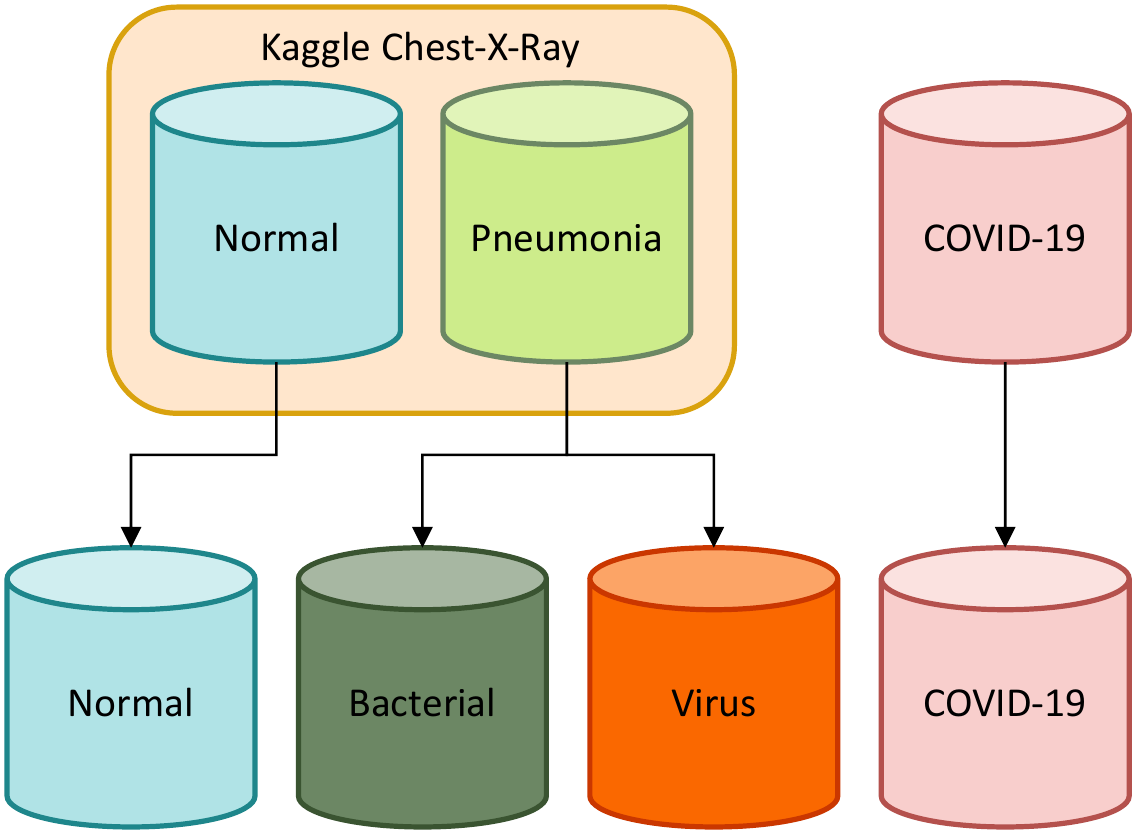}
	\caption{Dataset subdivision scheme.}
    \label{fig:dataset-split}
\end{figure}

\subsection{MobileNet v2}\label{par:mobile-netv2}
An earlier network devised for mobile devices is the MobileNet v2 \cite{sandler2018mobilenetv2}. This model implements a fully convolution layer followed by 19 residual bottleneck layers that extend the depth-wise separable convolution designed for the MobileNet \cite{howard2017mobilenets}. In detail, this operation applies convolutions on the input channels in a disjoint fashion, then merges the resulting outputs via 1$\times$1 point-wise convolution. In MobileNet v2, this operation is further refined through an inverted residual block that expands, performs a depth-wise separable convolution, then compresses the output to filter features; preventing overfitting and reducing the model parameter number. Finally, similarly to the MnasNet, the MobileNet v2 classifies a feature vector with dimension 1280 via a single dense layer.

\subsection{MobileNet v3}\label{par:mobile-netv3}
A further advancement for mobile networks is also provided by MobileNet v3 \cite{howard2019searching}, where a feature vector of size 1280 classified by a single dense layer is extracted by inverted residual blocks with depth-wise separable convolutions, i.e., MobileNet v2 blocks, that are improved through the squeeze-and-excitation technique \cite{hu2018squeeze}. Specifically, the latter explicitly models interdependencies between channels by recalibrating channel-wise feature responses through adaptive weights associated to each feature map. Furthermore, AutoML \cite{thornton2013auto} is also employed to find the best neural network architecture using the extended inverted residual blocks mentioned above.

\subsection{ResNet50}
An effective architecture focusing on the extracted features is the ResNet50 \cite{he2016deep} which, as the name implies, contains 50 convolutional layers. In more detail, this architecture applies a first convolution and a max pooling operation to consistently reduce the input size. Subsequently, residual blocks are implemented to reach the desired size of 50 convolutional layers that generate a feature vector of dimension 2048, classified by a single dense layer. Moreover, a key component of this architecture lies in its residual blocks, which leverage residual connections. The latter define an identity shortcut that skips 3 layers and ensures input data is feed-forwarded throughout the model, enabling for a higher information retention.

\subsection{ResNeXt}
The ResNeXt \cite{xie2017aggregated} architecture extends the ResNet residual blocks with a split-transform-merge paradigm similar to the GoogLeNet inception module. Differently from the latter, that concatenates depth-wise feature maps generated using different kernel sizes, ResNeXt merges the different paths by summation. What is more, each path will employ identical kernels for its convolutions, effectively designing an architecture by repeating a building block with a specific set of transformations; resulting in a simple and homogeneous design and reducing the number of hyper-parameters to be set. Finally, this architecture, similarly to the ResNet50, classifies a feature vector of size 2048 using a single dense layer.

\subsection{ShuffleNet}
The ShuffleNet model \cite{zhang2018shufflenet} is another network optimized for mobile devices that requires low computational resources. To reduce the latter, generally measured in Mega Floating-point Operations per Second (MFLOPs), this model defines two operations that reduce the computational costs while still preserving accuracy, i.e., point-wise group convolution and channel shuffle. These two procedures rearrange feature maps channels to increase the architecture abstraction level, achieving high performances on mobile devices (e.g., 18x faster execution with respect to an AlexNet on ARM-based devices, for similar accuracy) even though the structure contains 50 layers organized into $4$ stages that apply the aforementioned mechanisms. Notice that similarly to other architectures, ShuffleNet implements a single dense layer to classify a feature vector containing 1024 elements to reduce the amount of parameters in the classifier.

\subsection{SqueezeNet}
Differently from other networks, SqueezeNet \cite{iandola2016squeezenet} focuses on significantly reducing the model size, resulting in a compact CNN with 18 convolutional layers generating a feature vector with dimension 1000, classified by a single dense layer that has similar performances to AlexNet on the ImageNet dataset \cite{krizhevsky2012imagenet}, but with 50x less parameters. To obtain these results, SqueezeNet authors describe Fire layers implementing an alternating squeeze-expand strategy, where parameters are constrained by reducing the convolution filter size. In particular, 1$\times$1 as well as 1$\times$1 and 3$\times$3 kernels are used in the squeeze and expand components, respectively. As a result, the parameter number reduction allows for more feature maps to be generated in deeper layers to enhance the model accuracy.

\begin{table}[t]
    \centering
    \begin{tabular}{lc}
		\toprule
		Hyperparameter & Value \\
		\midrule
		Max Epochs     & $40$ \\
		Optimizer      & Adam \\
		Learning Rate  & 1$\mathrm{e}$-3 \\
		Batch size     & $100$ \\
		Loss function  & Categorical Cross-Entropy \\
		Model Selection & High Accuracy \\
		\bottomrule
    \end{tabular}
    \caption{Model training hyperparameter configuration.}
    \label{tab:hparams}
\end{table}

\subsection{VGG16}
One of the first networks to include a deeper architecture is the OxfordNet model, also known as VGG16 \cite{simonyan2014very}. The Oxford Visual Geometry Group achieved remarkable performances by stacking up to $13$ convolutional layers, interleaved by $5$ max pooling operations applied at the second, fourth, seventh, tenth, and thirteenth convolutions to reduce the input size. Through this structure, a vector of meaningful features with size 25088 is extracted from the input image and fed to a classifier composed of three dense layers that discriminates between the ImageNet $1000$ classes. Moreover, the VGG16 is one of the first models applying smaller kernels in its convolutions, i.e., with a shape of 3$\times$3, used to capture details in extremely small receptive fields.

\subsection{Wide ResNet}
A different approach to typical deep neural network design is explored by the Wide ResNet model \cite{zagoruyko2016wide}. In this architecture, ResNet residual blocks, which were otherwise devised to lessen the number of parameters of deep neural networks by allowing for deep, thin and effective structures, are modified to reduce their depths and increase their width. In particular, this is achieved by limiting the amount of layers in residual blocks, while employing bigger kernels and generating more feature maps per convolution. A strategy that enabled the Wide ResNet with its 16-layer structure, followed by a single dense layer classifier taking as input a feature vector of 2048 elements, to outperform other deeper models on the ImageNet Large Scale Visual Recognition challenge \cite{russakovsky2015imagenet}.

\begin{table}[t]
    \centering
        \begin{tabular}{l|cc|r}
            \toprule
            Model & Training time & Test time & Params\\
            \midrule
            AlexNet & 17m & 19s & 58M \\
            DenseNet & 48m & 19s & 27M \\
            GoogleNet & 15m & 19s & 5M \\
            MnasNet & 37m & 18s & 3M \\
            MobileNet v2 & 37m & 18s & 2M \\
            MobileNet v3 & 39m & 18s & 3M \\
            ResNet50 & 36m & 19s & 24M \\
            ResNext & 43m & 20s & 23M \\
            ShuffleNet & 50m & 19s & 1M \\
            SqueezeNet & 17m & 18s & 0.7M \\
            VGG16 & 42m & 22s & 135M \\
            Wide ResNet50 & 47m & 21s & 67M \\
            \bottomrule
        \end{tabular}%
    \caption{Training and test time comparison expressed in (m)inutes and (s)econds, respectively. Lower values correspond to a better performance.}
    \label{tab:training-time}
\end{table}

\section{Experimental Results and Discussion}\label{sec:experimental_results}
Extensive experiments were carried out to assess the transfer learning capabilities of all networks. The dataset employed for evaluation and its merging procedure are described in Section~\ref{subsec:dataset}. Implementation details and the chosen testing protocol are instead presented in Section~\ref{subsec:implementation_details}. Finally, quantitative and qualitative experimental results are shown and discussed in Section~\ref{subsec:performance_evaluation}.

\begingroup
\setlength{\tabcolsep}{3pt}
\begin{table}[t]
    \centering
    \resizebox{\columnwidth}{!}{
        \begin{tabular}{l|cccc|r}
            \toprule
            Model & Prec.\% & Sens.\% & Spec.\% & F1\% & \#Params \\
            \midrule
            AlexNet & 80.41 & 78.45 & 92.82 & 77.73 & 58M \\
            DenseNet & 84.70 & 83.01 & 94.34 & 84.00 & 27M \\
            GoogleNet & 83.09 & 80.52 & 93.51 & 81.33 & 5M \\
            MnasNet & 69.50 & 56.49 & 85.50 & 54.60 & 3M \\
            MobileNet v2 & 83.78 & 81.91 & 93.97 & 82.48 & 2M \\
            MobileNet v3 & \textbf{84.92} & \textbf{83.43} & \textbf{94.48} & \textbf{84.36} & 3M \\
            ResNet50 & 80.58 & 76.38 & 92.13 & 77.49 & 24M \\
            ResNext & 79.55 & 77.49 & 92.50 & 78.50 & 23M \\
            ShuffleNet & 83.19 & 78.04 & 92.68 & 79.33 & 1M \\
            SqueezeNet & 20.73 & 27.49 & 75.83 & 19.70 & 0.7M \\
            VGG16 & 77.39 & 74.72 & 91.57 & 75.35 & 135M  \\
            Wide ResNet50 & 79.83 & 74.45 & 91.48 & 76.23 & 67M \\
            \bottomrule
        \end{tabular}%
    }
    \caption{Performance comparison computed on the test set for precision, sensitivity, specificity, and F1-score metrics. All models are trained using 100\% of the training set.}
    \label{tab:results}
\end{table}
\endgroup

\begingroup
\setlength{\tabcolsep}{3pt}
\begin{table}[t]
    \centering
    \resizebox{\columnwidth}{!}{
        \begin{tabular}{l|cccc|r}
            \toprule
            Model & Prec.\% & Sens.\% & Spec.\% & F1\% & \#Params \\
            \midrule
            AlexNet & 79.58 & 77.07 & 92.36 & 76.36 & 58M \\
            DenseNet & \textbf{83.97} & \textbf{79.97} & \textbf{93.32} & \textbf{81.07} & 27M \\
            GoogleNet & 80.78 & 76.80 & 92.27 & 77.41 & 5M \\
            MnasNet & 68.93 & 51.66 & 83.89 & 48.06 & 3M \\
            MobileNet v2 & 82.16 & 78.31 & 92.77 & 80.16 & 2M \\
            MobileNet v3 & 82.67 & 79.00 & 93.00 & 79.85 & 3M \\
            ResNet50 & 78.46 & 71.13 & 90.38 & 72.46 & 24M \\
            ResNext & 77.74 & 70.44 & 90.15 & 71.30 & 23M \\
            ShuffleNet & 80.56 & 72.51 & 90.84 & 73.19 & 1M \\
            SqueezeNet & 11.17 & 33.42 & 87.81 & 12.53 & 0.7M \\
            VGG16 & 70.62 & 67.96 & 89.32 & 67.00 & 135M \\
            Wide ResNet50 & 75.27 & 70.30 & 90.10 & 70.90 & 67M \\
            \bottomrule
        \end{tabular}%
    }
    \caption{Performance comparison computed on the test set for precision, sensitivity, specificity, and F1-score metrics. All models are trained using 50\% of the training set.}
    \label{tab:results-50}
\end{table}
\endgroup

\subsection{Dataset}\label{subsec:dataset}
To explore transfer learning capabilities on pneumonia classification of chest-x-rays images, with particular attention to the SARS-CoV-2 infection, we performed experiments on a dataset that was specifically designed to contain data from different public collections addressing the pneumonia classification task. Notice that the merged collection has heterogeneous chest-x-ray dimensions due to its data deriving from distinct research groups. However, when fed to the selected models, the images are resized to a shape of 224$\times$224 to match the required input size of the architectures. In detail, the merged dataset contains four categories, namely: normal (i.e., healthy), and pneumonia deriving from bacteria, generic viruses, or COVID-19, as illustrated in Fig.~\ref{fig:dataset-split}.

Regarding the first three classes, they are taken from a well-known Kaggle pneumonia challenge, and split into training, development, and test sets in accordance with \cite{kermany2018identifying}. This dataset comprises more than 5000 (anterior-posterior) chest-x-rays images that were selected from retrospective cohorts of pediatric patients with an age between one and five years old during the pre-COVID-19 era. In detail, the images were collected by the Guangzhou Women and Children’s Medical Center, and they were graded by at least two expert physicians. However, the original dataset is annotated as a binary classification task, i.e., images either containing pneumonia symptoms or not. Nevertheless, thanks to its metadata it is possible to further split the images according to the illness source, i.e., either bacterial or virus, therefore resulting in $3$ of the aforementioned classes. 

Concerning chest-x-rays images showing a COVID-19 contagion, we further extend the training and test sets with 185 images by following the same arrangement proposed in \cite{minaee2020deep}. Moreover, since the latter does not have any validation data, we also expand the corresponding set through the remarkable works of \cite{cohen2020covidProspective, cohen2020covid}, which assembled $\approx$400 chest-x-rays COVID-19 samples from different publications. In particular, these images were acquired by different institutions across the world and include both male and female patients covering all ages up to 80 years old. From this collection, 289 images were retained to ensure there were non-overlapping images with respect to sets defined by \cite{minaee2020deep}.

Summarizing, the final dataset, fully described in Table \ref{tab:dataset}, comprises the bacteria, virus, and normal classes as per \cite{kermany2018identifying}, while for COVID-19 it contains train and test sets from \cite{minaee2020deep}, and a validation set taken from \cite{cohen2020covid, cohen2020covidProspective}, for a total of 6330 chest-x-ray images.

\begingroup
\setlength{\tabcolsep}{3pt}
\begin{table}[t]
    \centering
    \resizebox{\columnwidth}{!}{
        \begin{tabular}{l|cccc|r}
            \toprule
            Model & Prec.\% & Sens.\% & Spec.\% & F1\% & \#Params \\
            \midrule
            AlexNet & 79.41 & 76.66 & 92.22 & 75.48 & 58M \\
            DenseNet & \textbf{81.20} & \textbf{77.21} & \textbf{92.40} & \textbf{78.00} & 27M \\
            GoogleNet & 79.92 & 75.69 & 91.90 & 75.12 & 5M \\
            MnasNet & 47.79 & 45.86 & 81.95 & 32.07 & 3M \\
            MobileNet v2 & 78.53 & 73.48 & 91.16 & 74.80 & 2M \\
            MobileNet v3 & 76.73 & 45.03 & 81.68 & 39.21 & 3M \\
            ResNet50 & 77.13 & 65.19 & 88.40 & 68.58 & 24M \\
            ResNext & 76.17 & 70.86 & 90.29 & 70.03 & 23M \\
            ShuffleNet & 76.56 & 67.13 & 89.04 & 61.98 & 1M \\
            SqueezeNet & 11.17 & 33.43 & 77.81 & 12.53 & 0.7M \\
            VGG16 & 69.57 & 62.43 & 87.48 & 62.34 & 135M \\
            Wide ResNet50 & 74.22 & 66.99 & 89.00 & 66.86 & 67M \\
            \bottomrule
        \end{tabular}%
    }
    \caption{Performance comparison computed on the test set for precision, sensitivity, specificity, and F1-score metrics. All models are trained using 20\% of the training set.}
    \label{tab:results-20}
\end{table}
\endgroup

\begingroup
\setlength{\tabcolsep}{3pt}
\begin{table}[t]
    \centering
    \resizebox{\columnwidth}{!}{
        \begin{tabular}{l|cccc|r}
            \toprule
            Model & Prec.\% & Sens.\% & Spec.\% & F1\% & \#Params \\
            \midrule
            AlexNet & 79.27 & 76.10 & 92.03 & 74.34 & 58M \\
            DenseNet & \textbf{81.26} & \textbf{76.93} & \textbf{92.31} & \textbf{77.96} & 27M \\
            GoogleNet & 78.92 & 71.41 & 90.47 & 70.96 & 5M \\
            MnasNet & 49.74 & 42.96 & 80.99 & 29.92 & 3M \\
            MobileNet v2 & 79.79 & 62.29 & 87.43 & 65.23 & 2M \\
            MobileNet v3 & 56.56 & 44.61 & 81.54 & 28.40 & 3M \\
            ResNet50 & 77.25 & 62.15 & 87.38 & 63.37 & 24M \\
            ResNext & 75.57 & 62.02 & 87.34 & 63.15 & 23M  \\
            ShuffleNet & 61.01 & 67.40 & 89.13 & 53.85 & 1M \\
            SqueezeNet & 11.22 & 33.29 & 77.76 & 12.55 & 0.7M  \\
            VGG16 & 68.16 & 58.29 & 89.31 & 59.84 & 135M  \\
            Wide ResNet50 & 76.51 & 72.51 & 90.84 & 70.44 & 67M \\
            \bottomrule
        \end{tabular}%
    }
    \caption{Performance comparison computed on the test set for precision, sensitivity, specificity, and F1-score metrics. All models are trained using 10\% of the training set.}
    \label{tab:results-10}
\end{table}
\endgroup

\subsection{Implementation Details}\label{subsec:implementation_details}
All experiments performed across the various architectures make use of a standard set of hyperparameters, summarized in Table \ref{tab:hparams}, to have a comparable environment across the models. In detail, each network was trained for up to $40$ epochs using the Adam optimizer, a learning rate of 1e-3, and a batch size of $100$, which allowed for a good sample mixture. Moreover, for each model, weights associated to the highest performing epoch, with respect to the development set accuracy, were selected as a final configuration to be used for inference. In addition, we also implemented the standard data augmentation strategy used for ImageNet which automatically handles input images with different dimensions by altering them as follows: random resize crop to a dimension of 256$\times$256, random rotation up to 15 degrees, color jitter, random horizontal flip, and center crop of size 224$\times$224 to match the network input size. Finally, several common classification metrics, i.e., precision, sensitivity, specificity, and F1-score were used to evaluate the networks performances. 
Notice that, being in a multiclass classification scenario, a weighted average of the aforementioned metrics was used to estimate the models performances so that the dataset distribution would also be taken into account.

Concerning the architectures implementation, we employed the PyTorch framework and PyTorch Lightning library, with a 16-bit floating-point precision to speed up the computation. Furthermore, ImageNet pre-trained models were imported from the TorchVision library to exploit the transfer learning paradigm. Finally, all tests were performed on the Google Colab infrastructure with an Intel CPU x86-64 architecture using 25GB of RAM, along an NVIDIA Tesla V100 with 16GB of VRAM. The required training and test time using this configuration are summarized in Table~\ref{tab:training-time}. Notice that training times are affected by the underlying architecture but are not strictly correlated to the number of available parameters, i.e., more parameters do not necessarily require longer times to be trained. Contrary to this phenomenon, test times over the entire test set, i.e., 725 images, are similar across all models, indicating that each network, independently from its performance, can analyze a chest-x-ray and provide a feedback to the clinician in roughly 3 ms.

\begin{figure}[t]
	\color{red}
	\centering
	\includegraphics[width=\columnwidth]{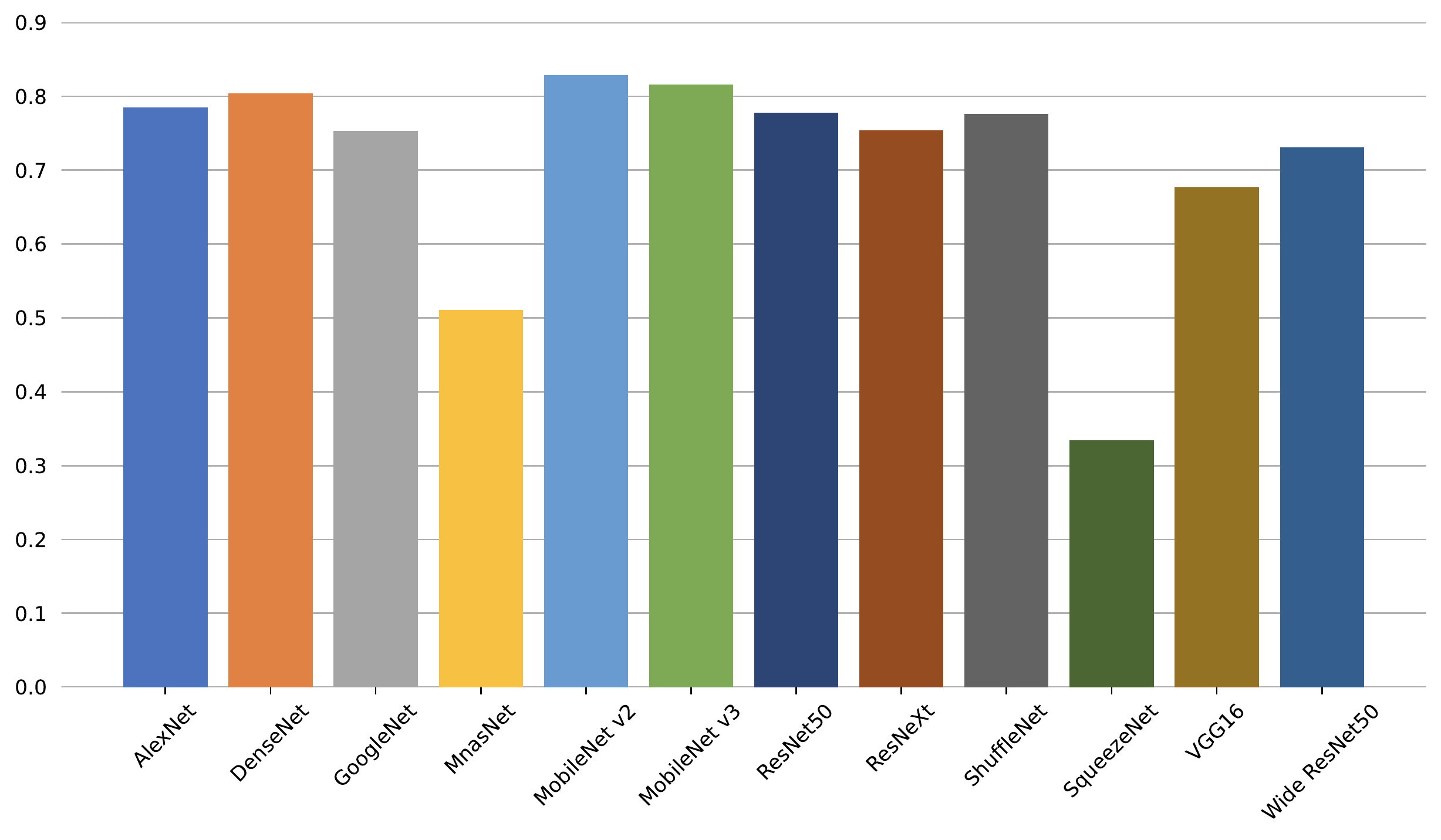}
	\caption{Accuracy computed on the test set.}
	\label{fig:test_acc}
\end{figure}

\begin{figure}[t]
    \centering
    \includegraphics[width=\columnwidth]{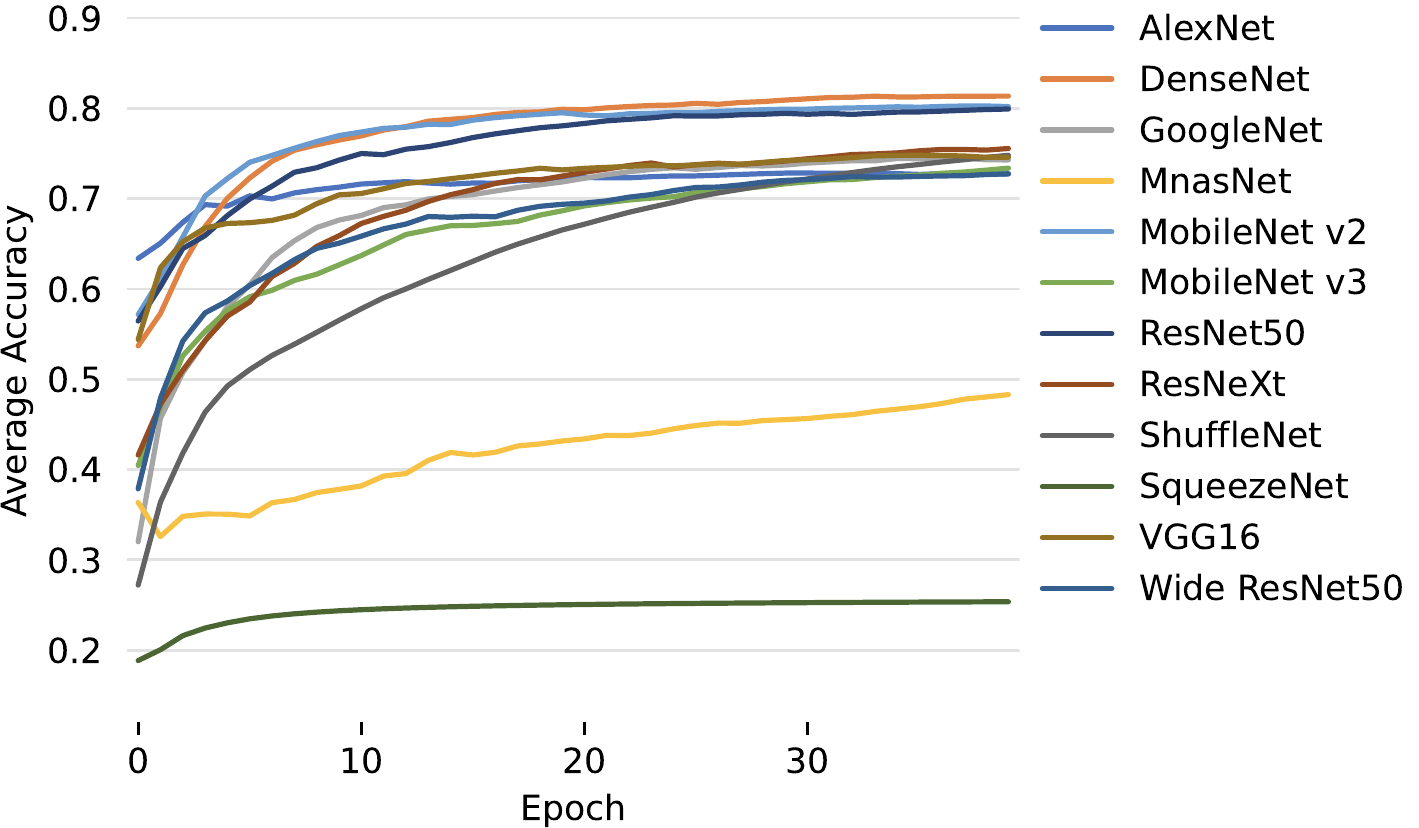}
    \caption{Average accuracy evolution during training.}
    \label{fig:acc_conv}
\end{figure}

\begin{figure}[t]
    \centering
    \includegraphics[width=\columnwidth]{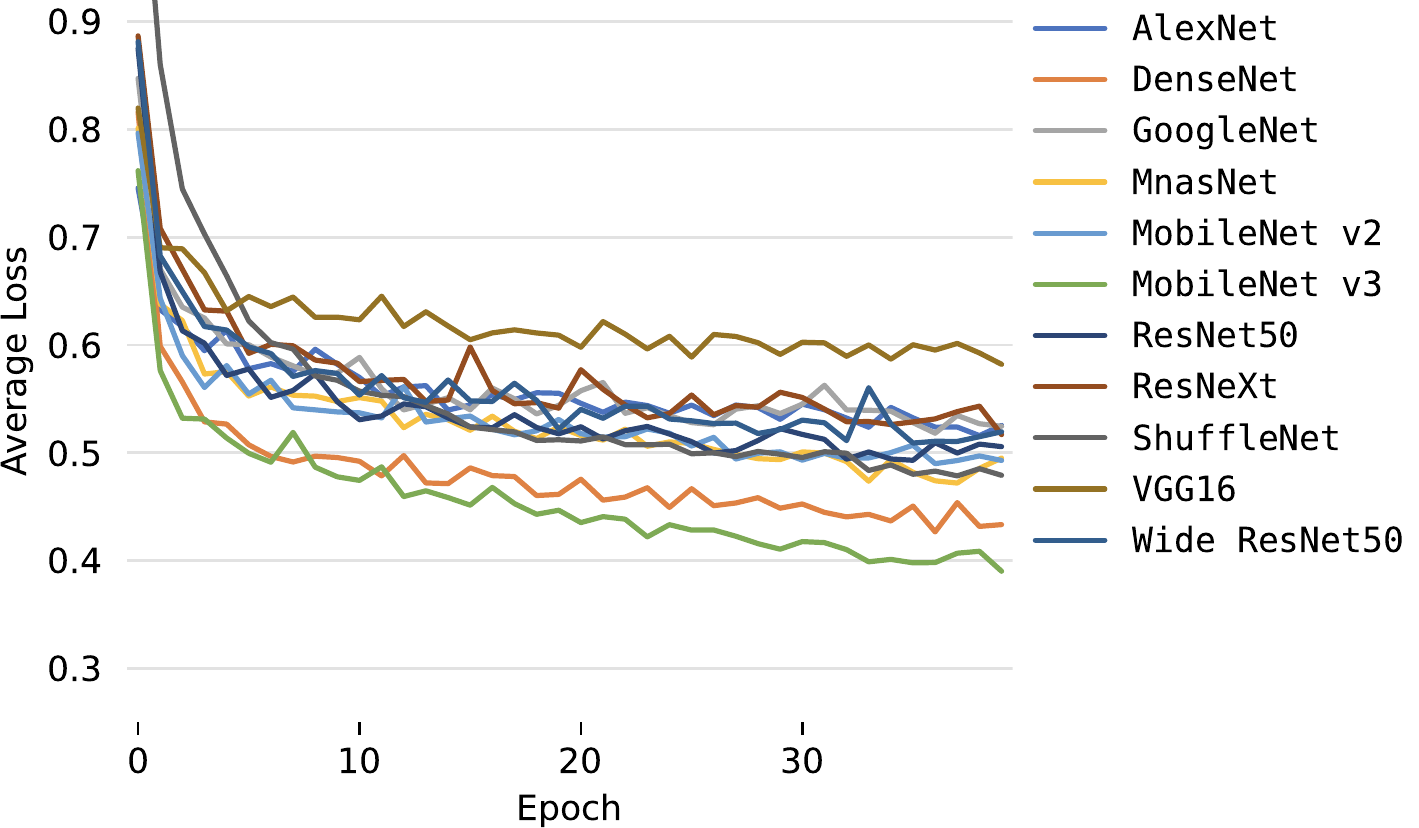}
    \caption{Average loss evolution during training.}
    \label{fig:loss_conv}
\end{figure}

\begin{figure*}[ht!]
    \centering
	\subfloat[]{\includegraphics[width=0.5\columnwidth]{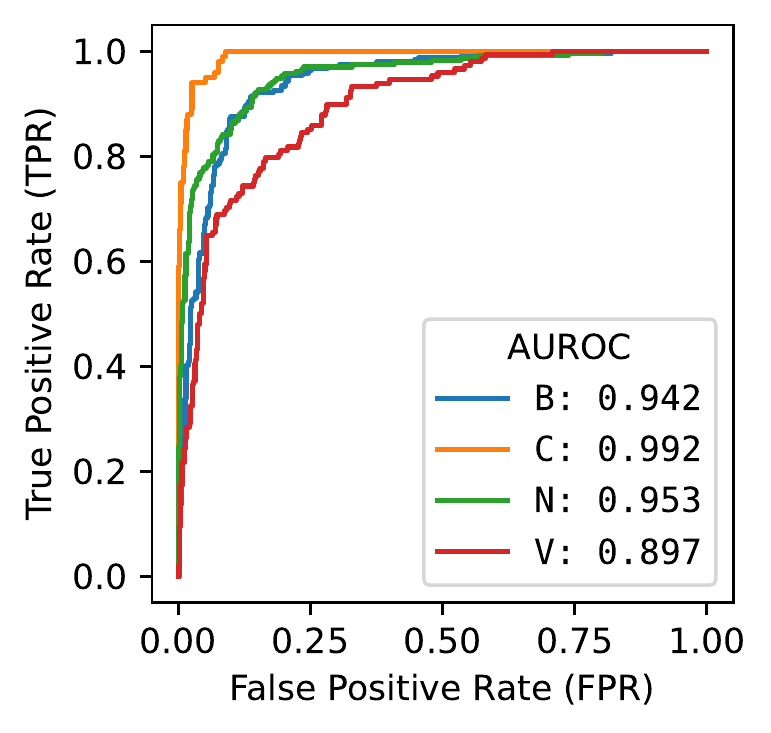}}
	\subfloat[]{\includegraphics[width=0.5\columnwidth]{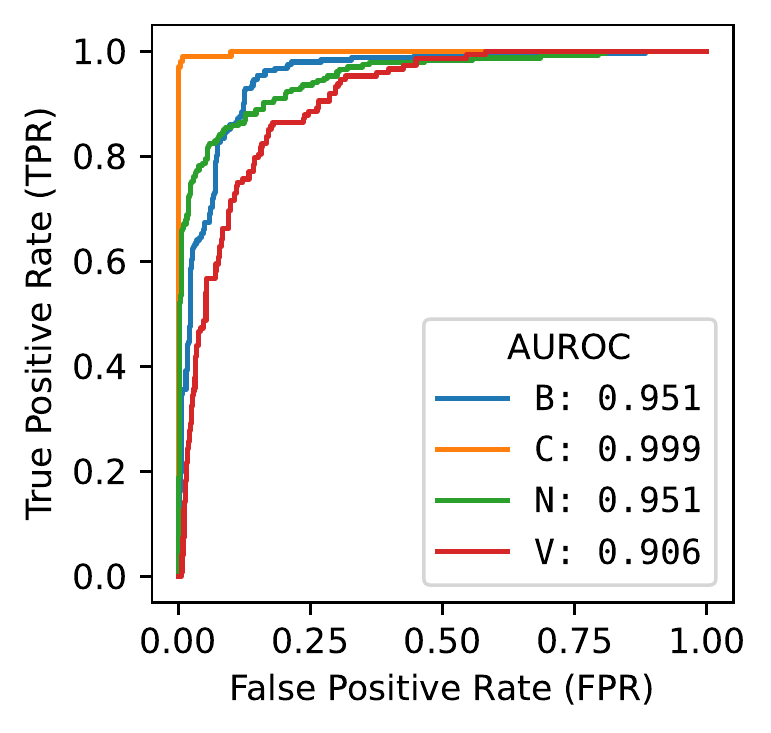}}
	\subfloat[]{\includegraphics[width=0.5\columnwidth]{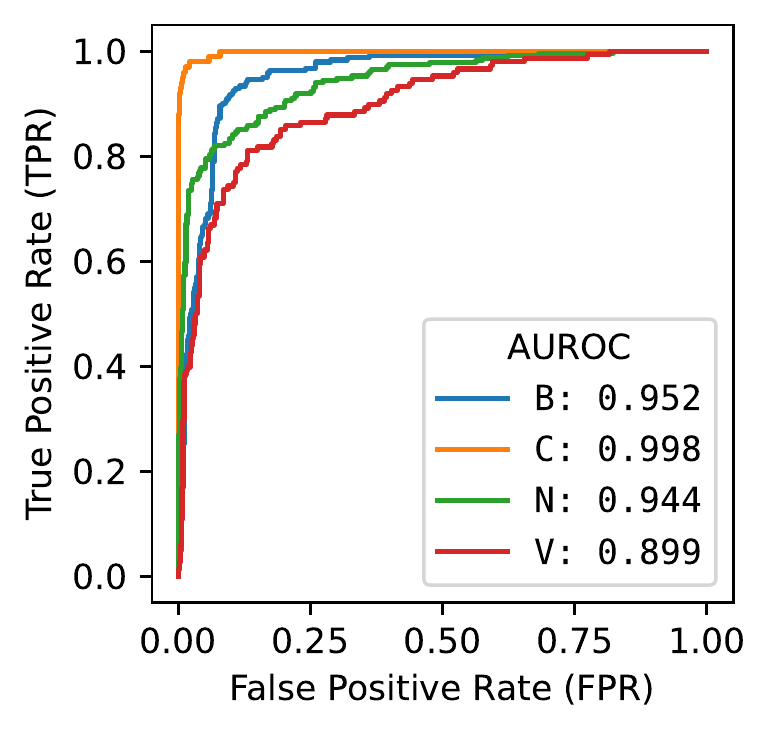}}
	\subfloat[]{\includegraphics[width=0.5\columnwidth]{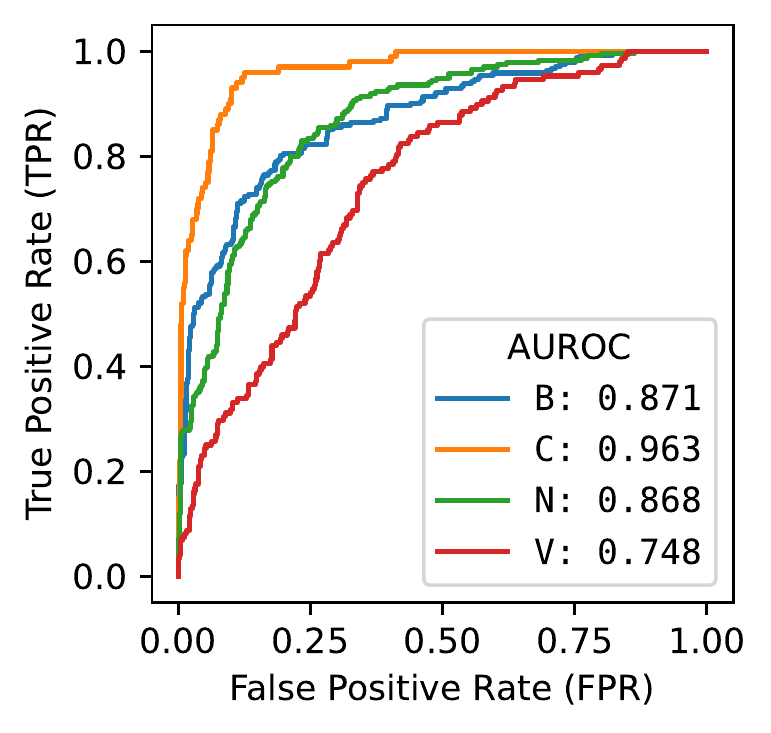}}
	\hfil
	\subfloat[]{\includegraphics[width=0.5\columnwidth]{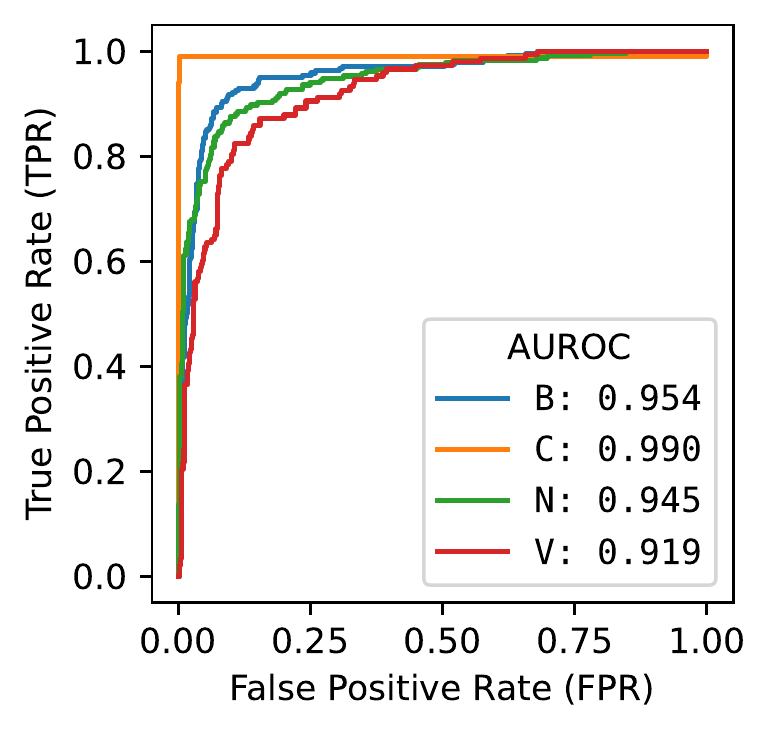}}
	\subfloat[]{\includegraphics[width=0.5\columnwidth]{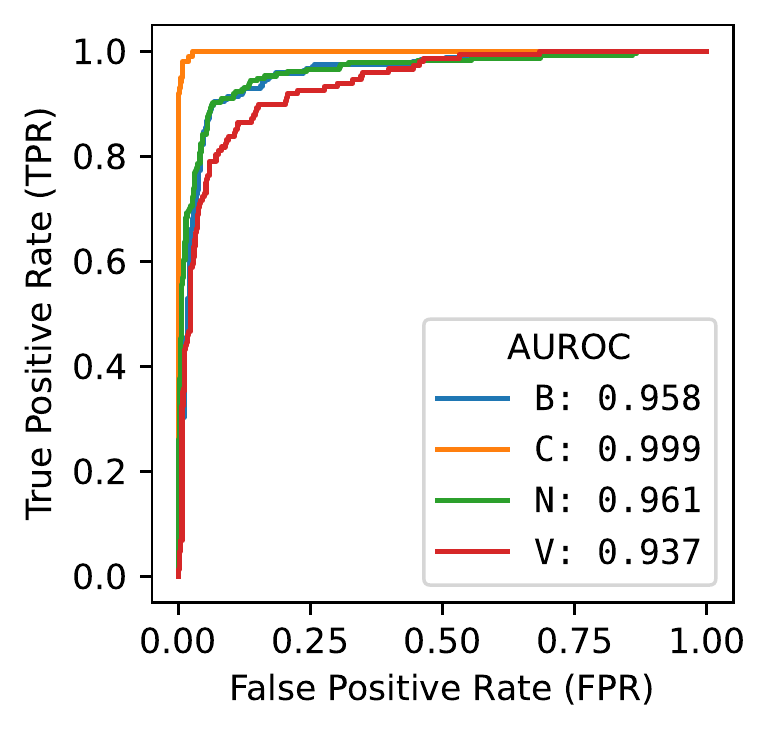}}
	\subfloat[]{\includegraphics[width=0.5\columnwidth]{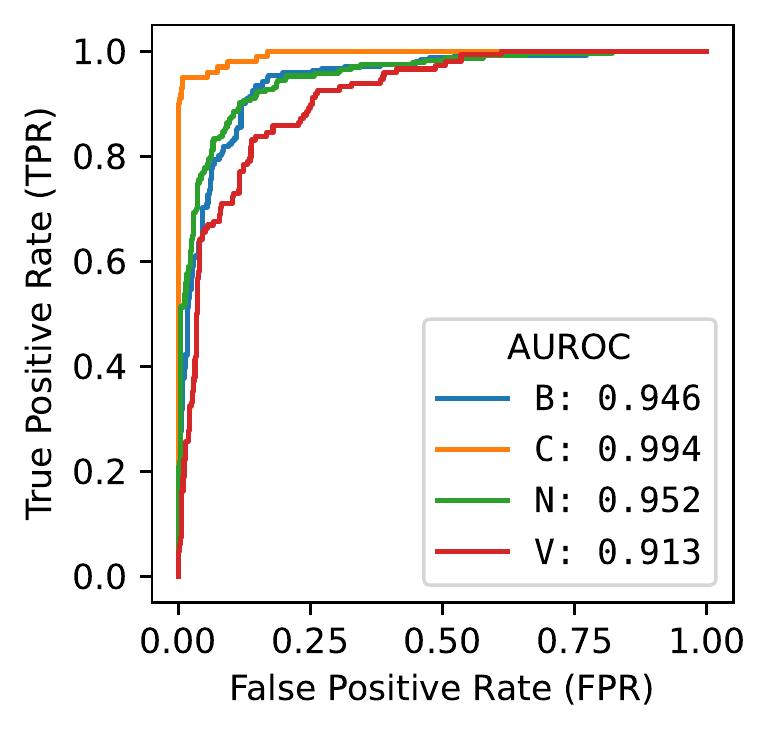}}
	\subfloat[]{\includegraphics[width=0.5\columnwidth]{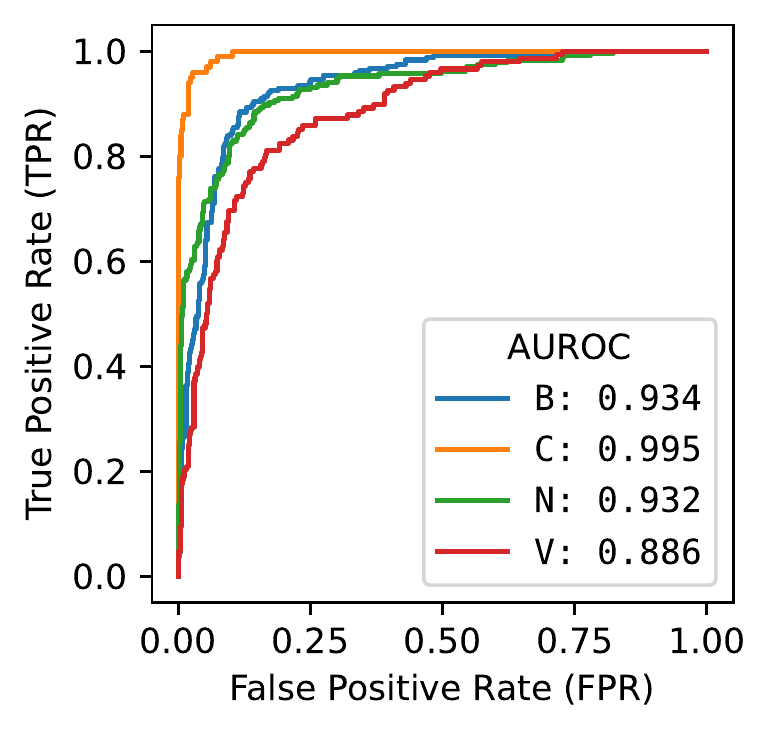}}
	\hfil
	\subfloat[]{\includegraphics[width=0.5\columnwidth]{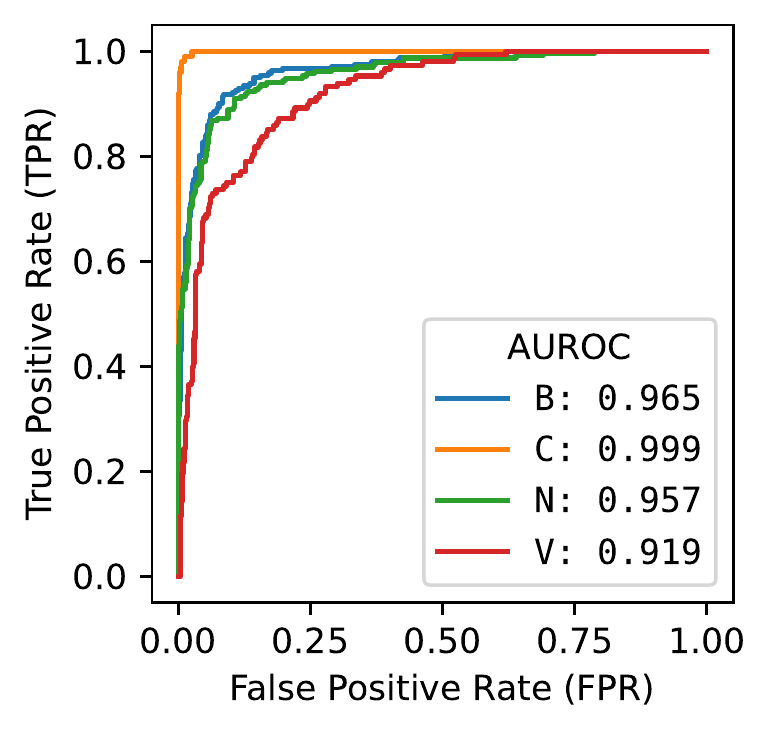}}
	\subfloat[]{\includegraphics[width=0.5\columnwidth]{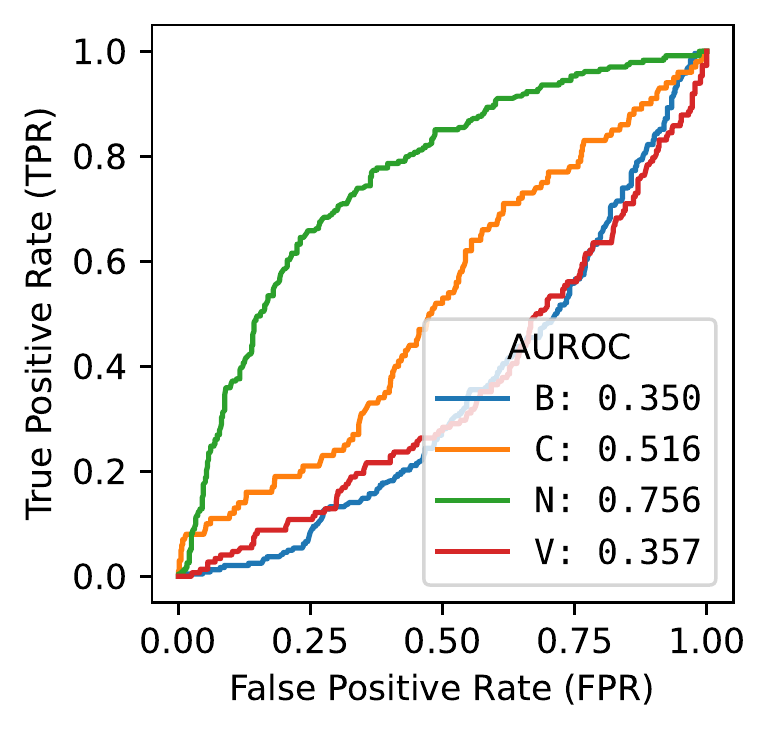}}
	\subfloat[]{\includegraphics[width=0.5\columnwidth]{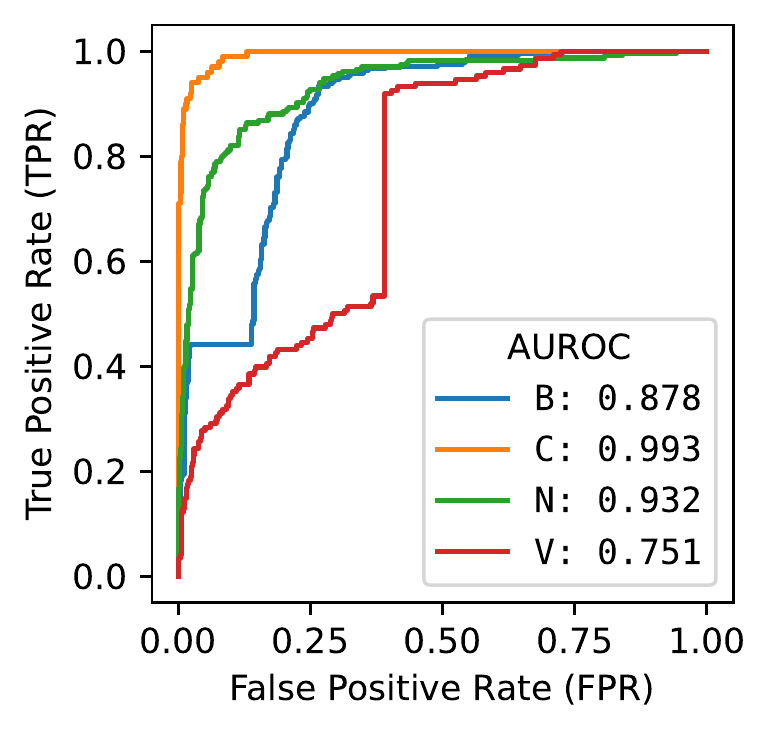}}
	\subfloat[]{\includegraphics[width=0.5\columnwidth]{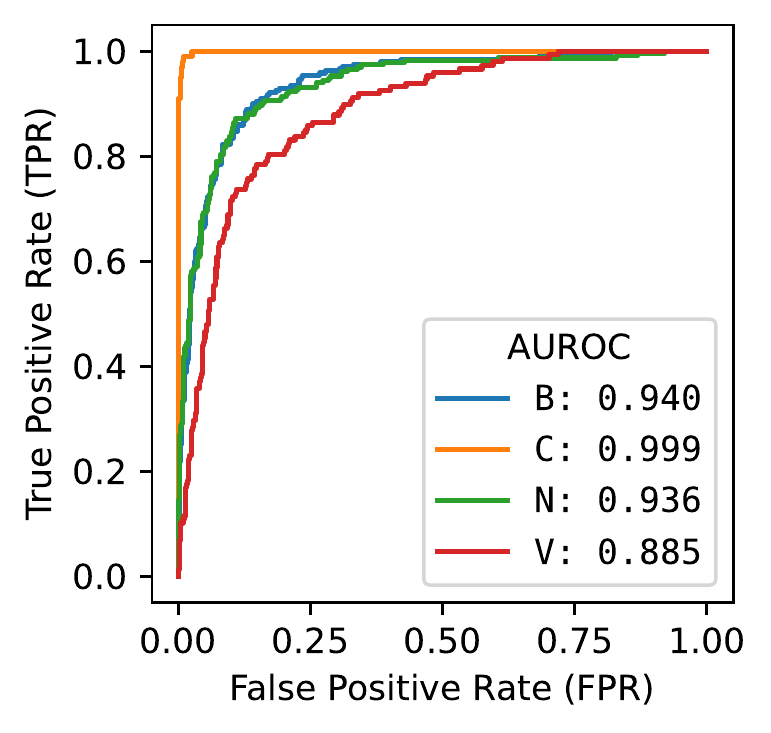}}
    \caption{AUROC metric computed for AlexNet (a), DenseNet (b), GoogleNet (c), MnasNet (d), MobileNet v2 (e), MobileNet v3 (Large) (f), ResNet50 (g), ResNext (h), ShuffleNet (i), SqueezeNet (j), VGG16 (k) and Wide ResNet50 (l), computed on the test set. Labels N, B, V, C correspond to normal, bacteria, virus and COVID-19 classes, respectively.}
    \label{fig:auroc}
\end{figure*}

\begin{figure*}[ht!]
    \centering
	\subfloat[]{\includegraphics[width=0.5\columnwidth]{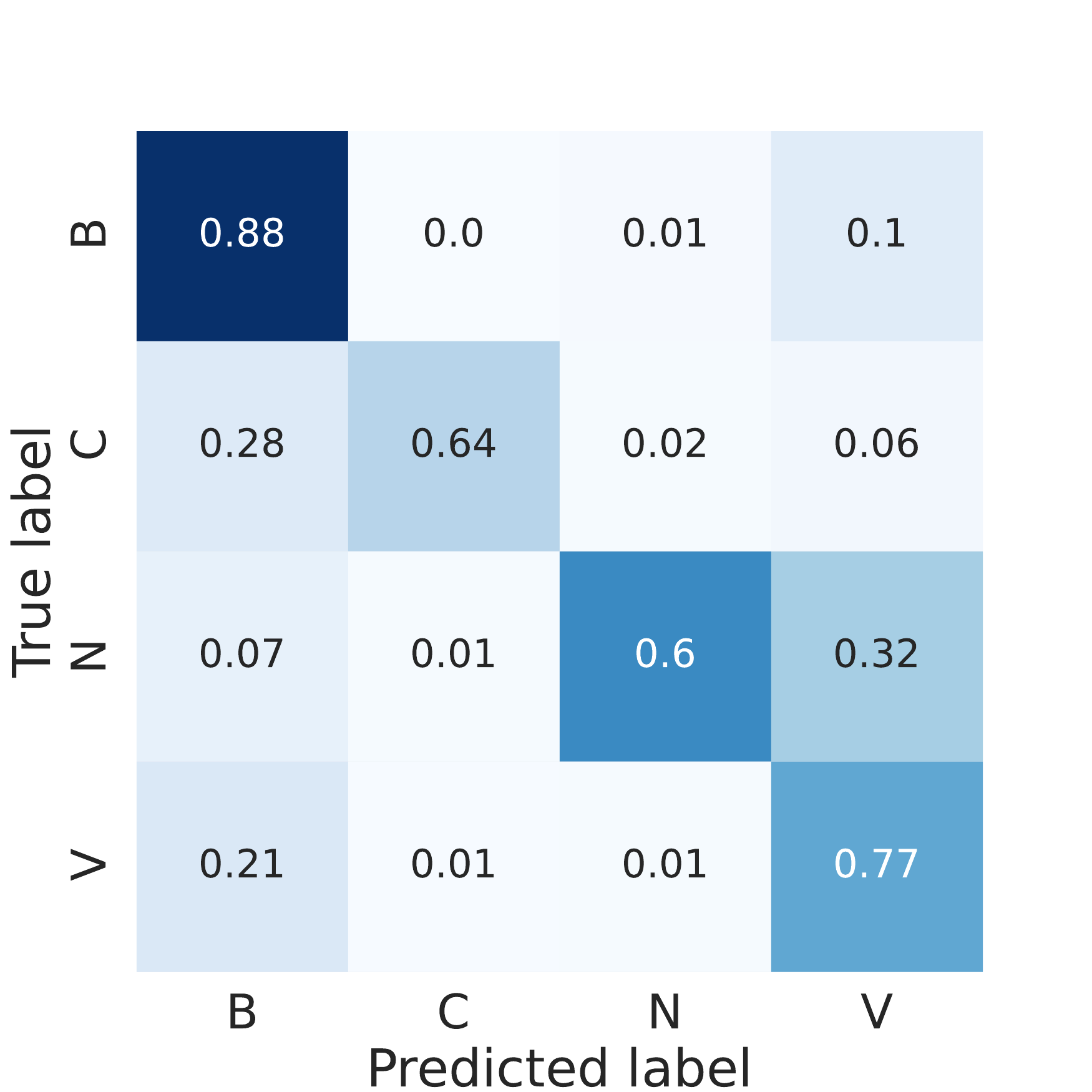}}
	\subfloat[]{\includegraphics[width=0.5\columnwidth]{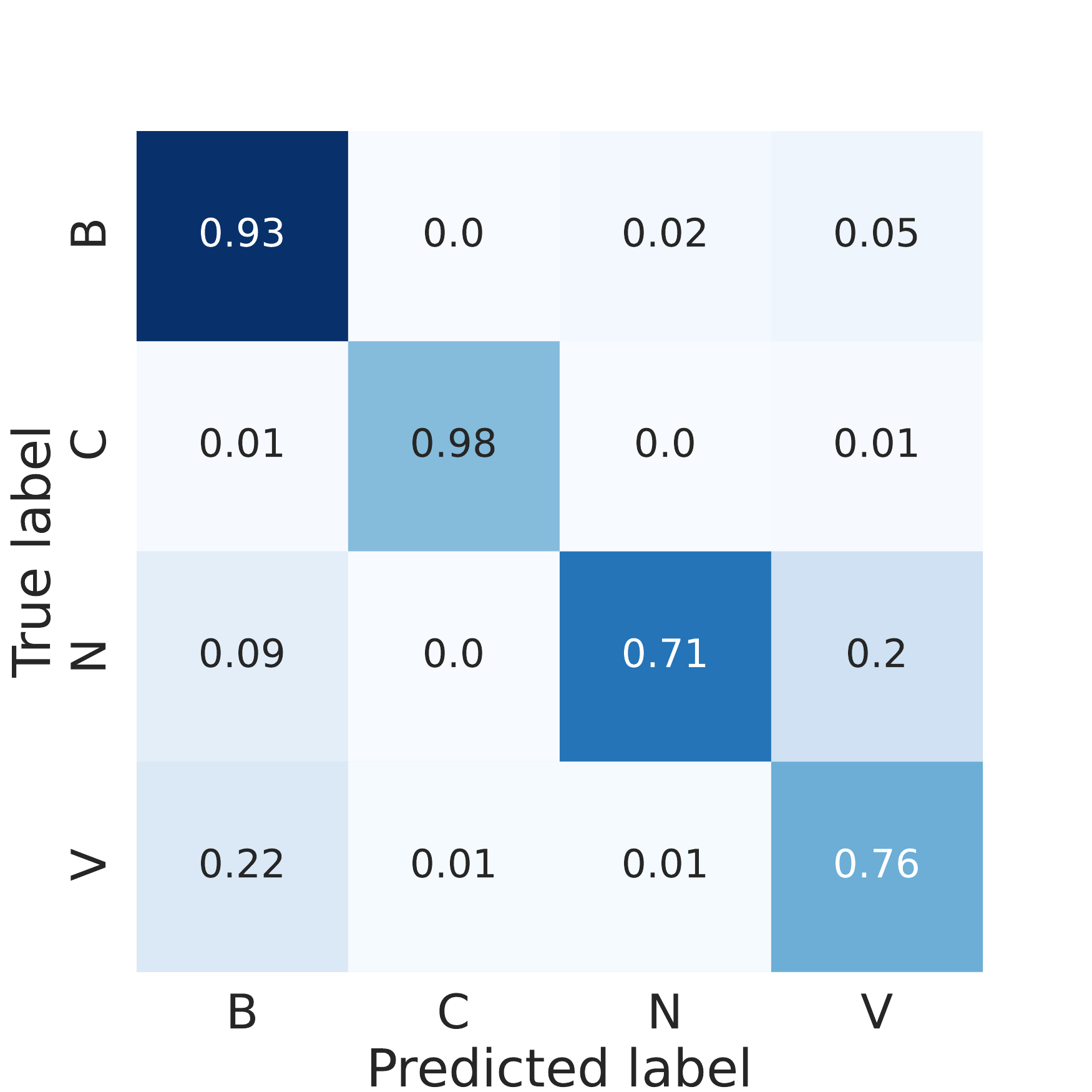}}
	\subfloat[]{\includegraphics[width=0.5\columnwidth]{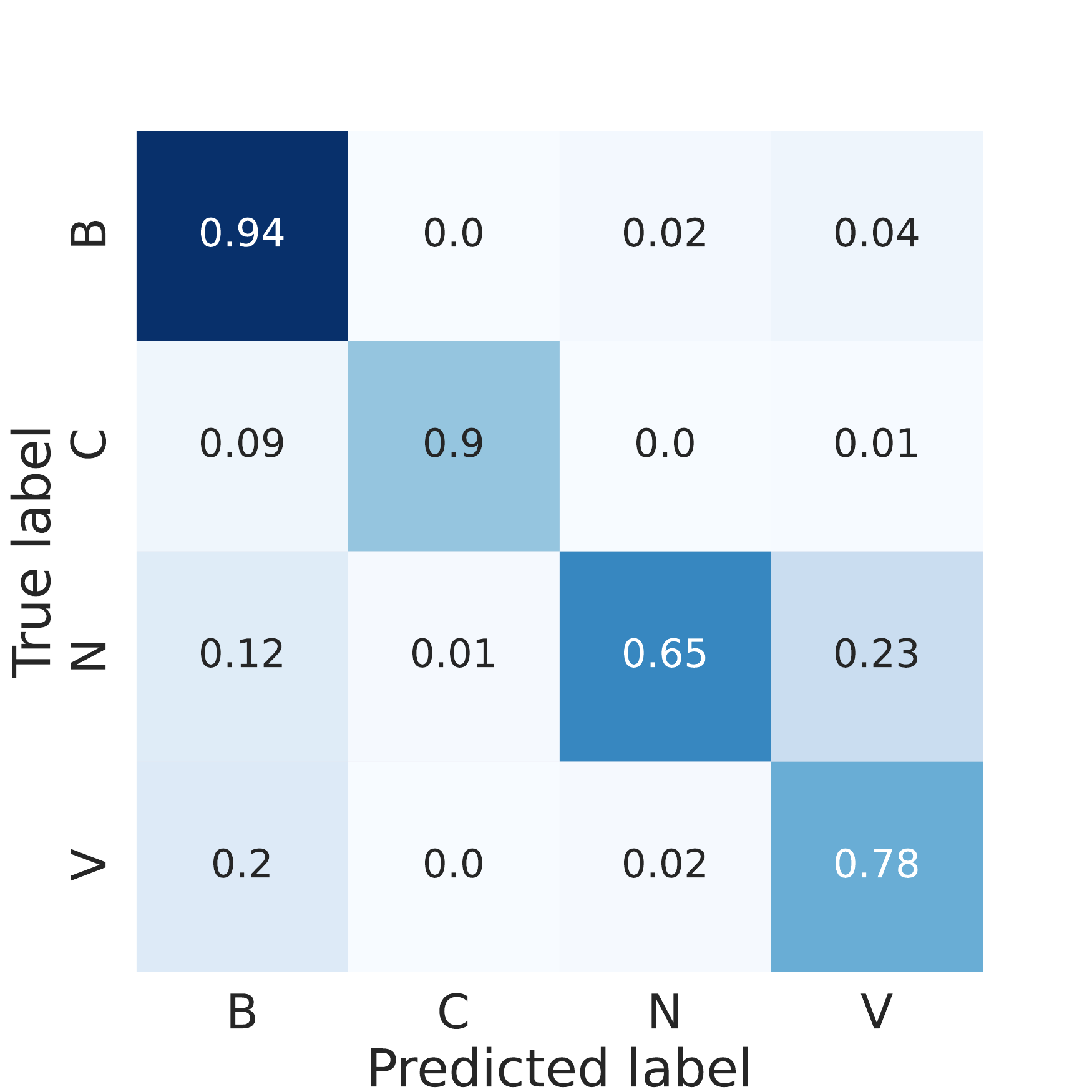}}
	\subfloat[]{\includegraphics[width=0.5\columnwidth]{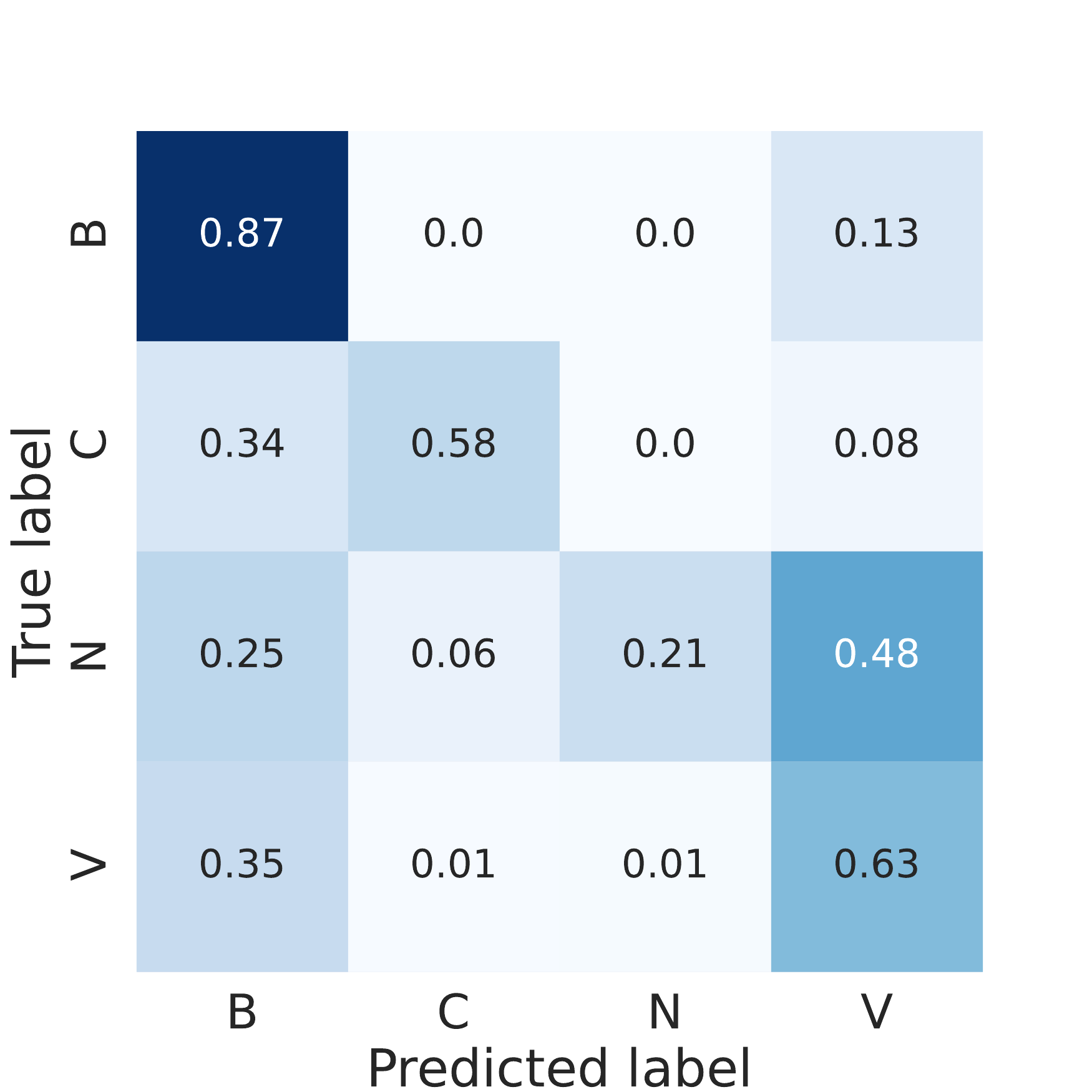}}
	\hfil
	\subfloat[]{\includegraphics[width=0.5\columnwidth]{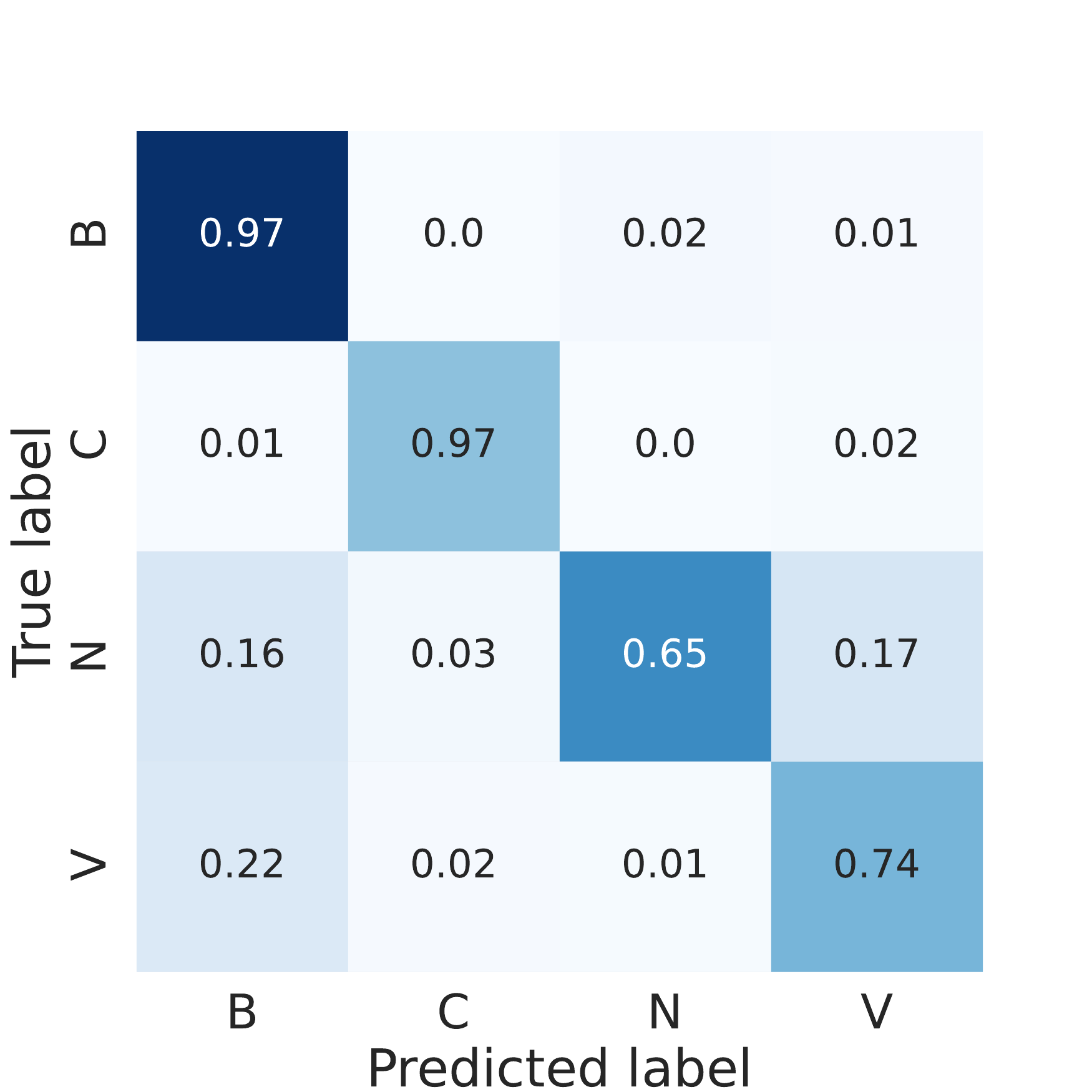}}
	\subfloat[]{\includegraphics[width=0.5\columnwidth]{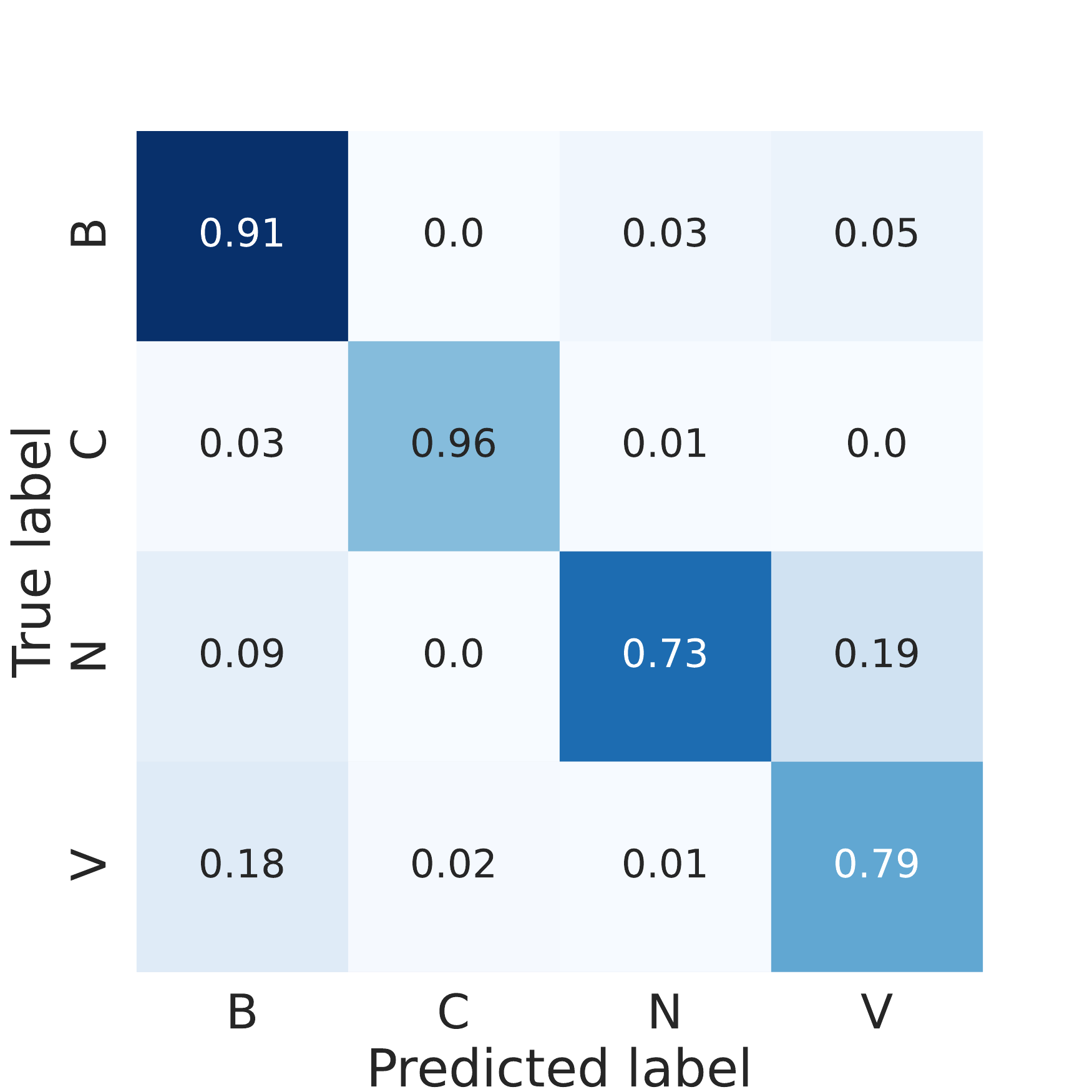}}
	\subfloat[]{\includegraphics[width=0.5\columnwidth]{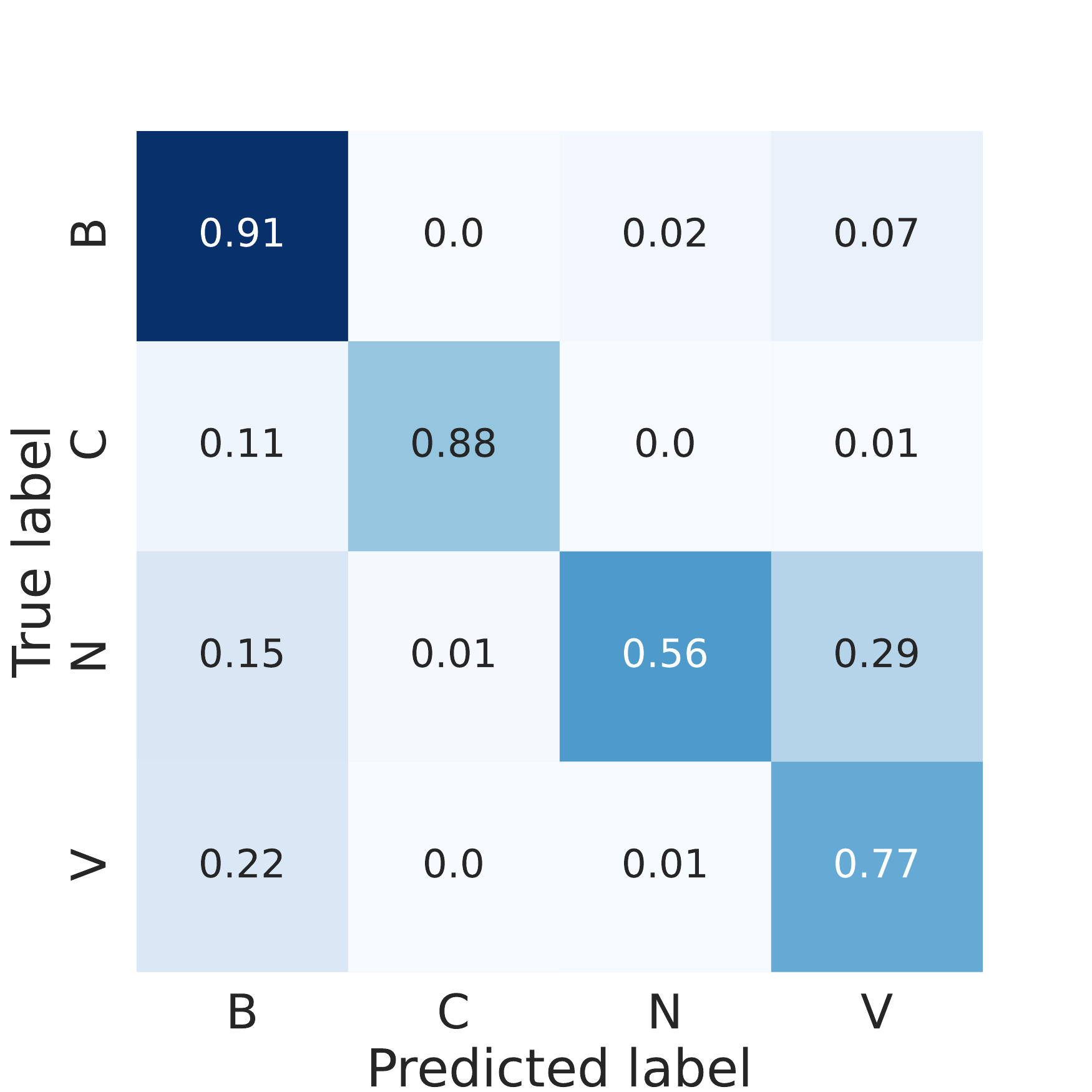}}
	\subfloat[]{\includegraphics[width=0.5\columnwidth]{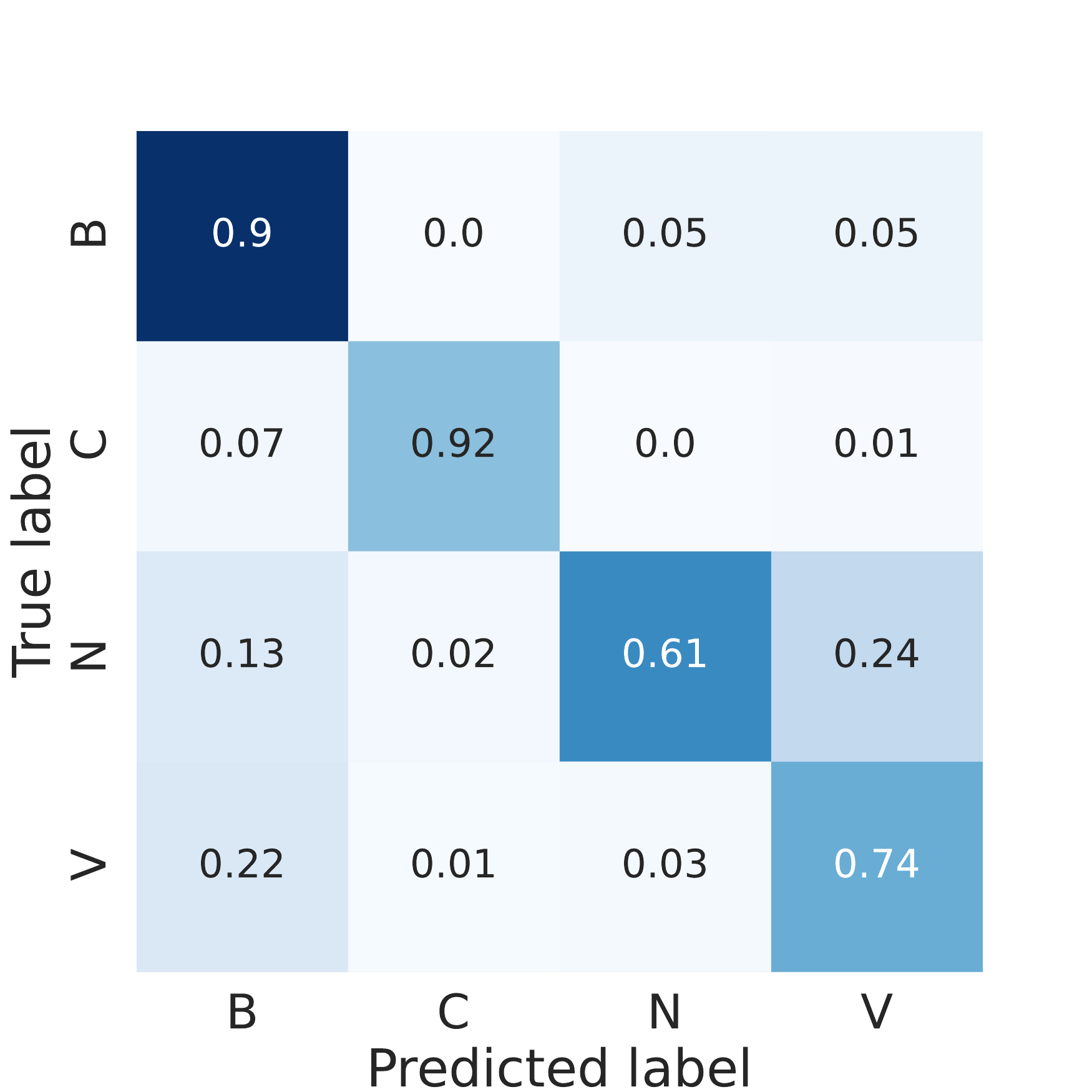}}
	\hfil
	\subfloat[]{\includegraphics[width=0.5\columnwidth]{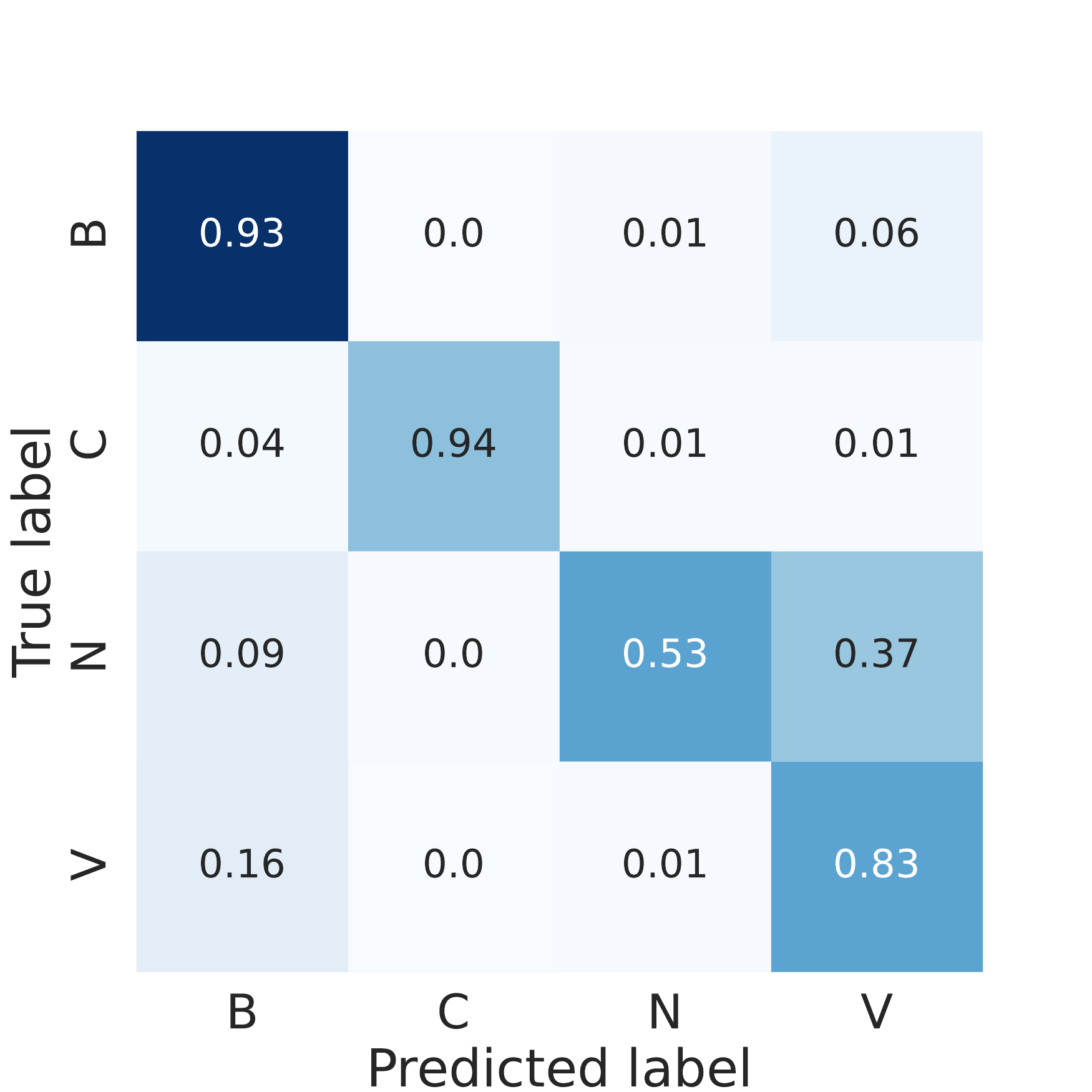}}
    \subfloat[]{\includegraphics[width=0.5\columnwidth]{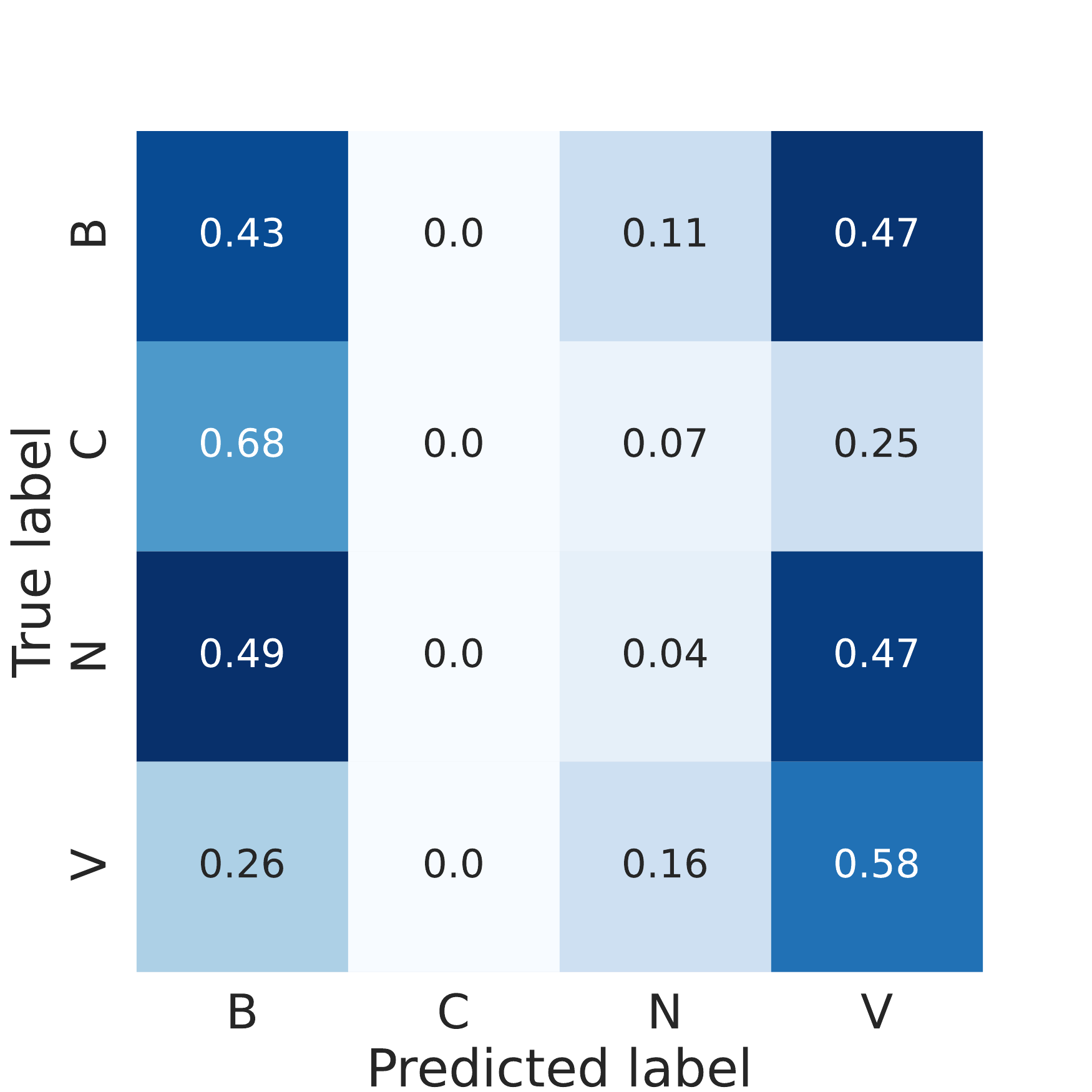}}
	\subfloat[]{\includegraphics[width=0.5\columnwidth]{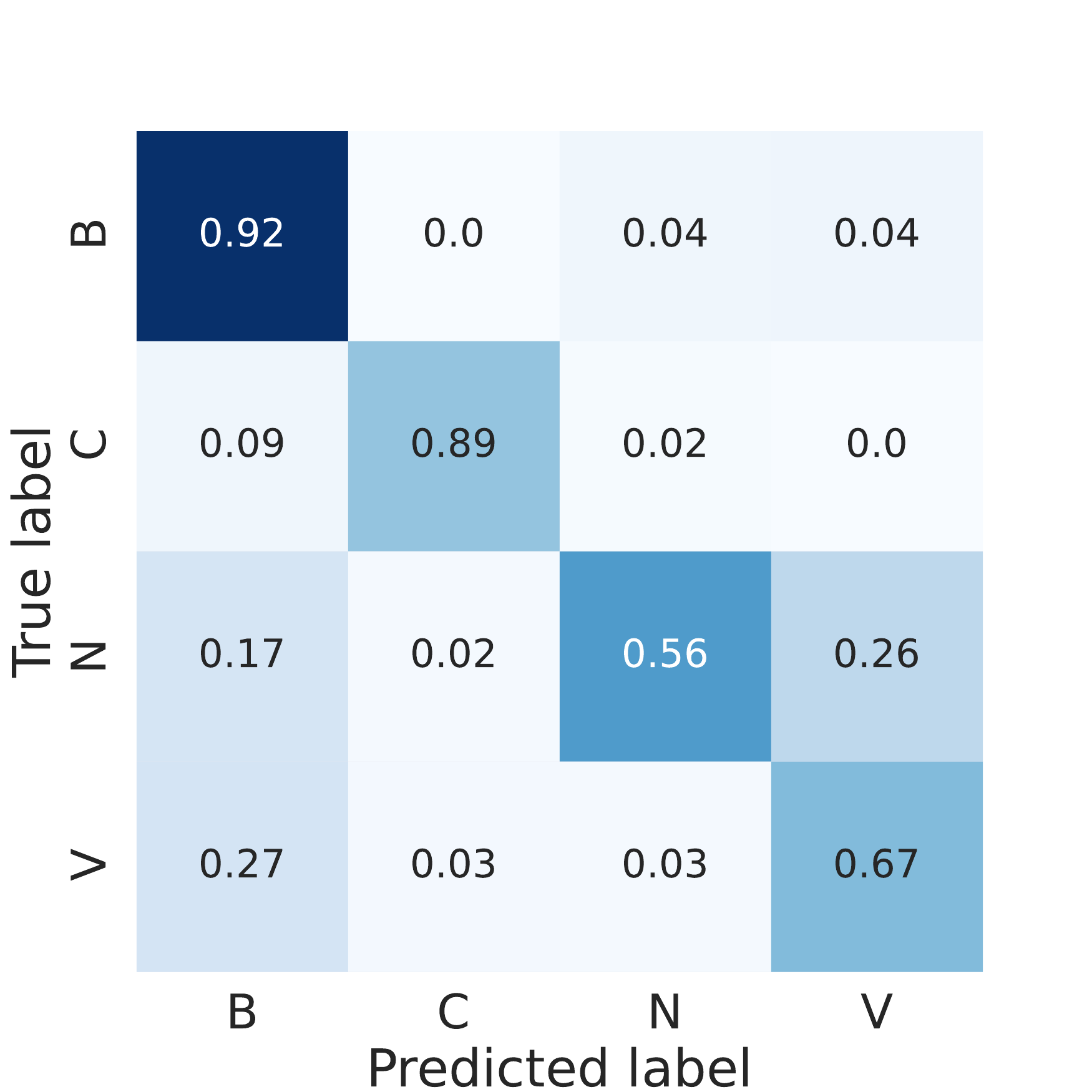}}
	\subfloat[]{\includegraphics[width=0.5\columnwidth]{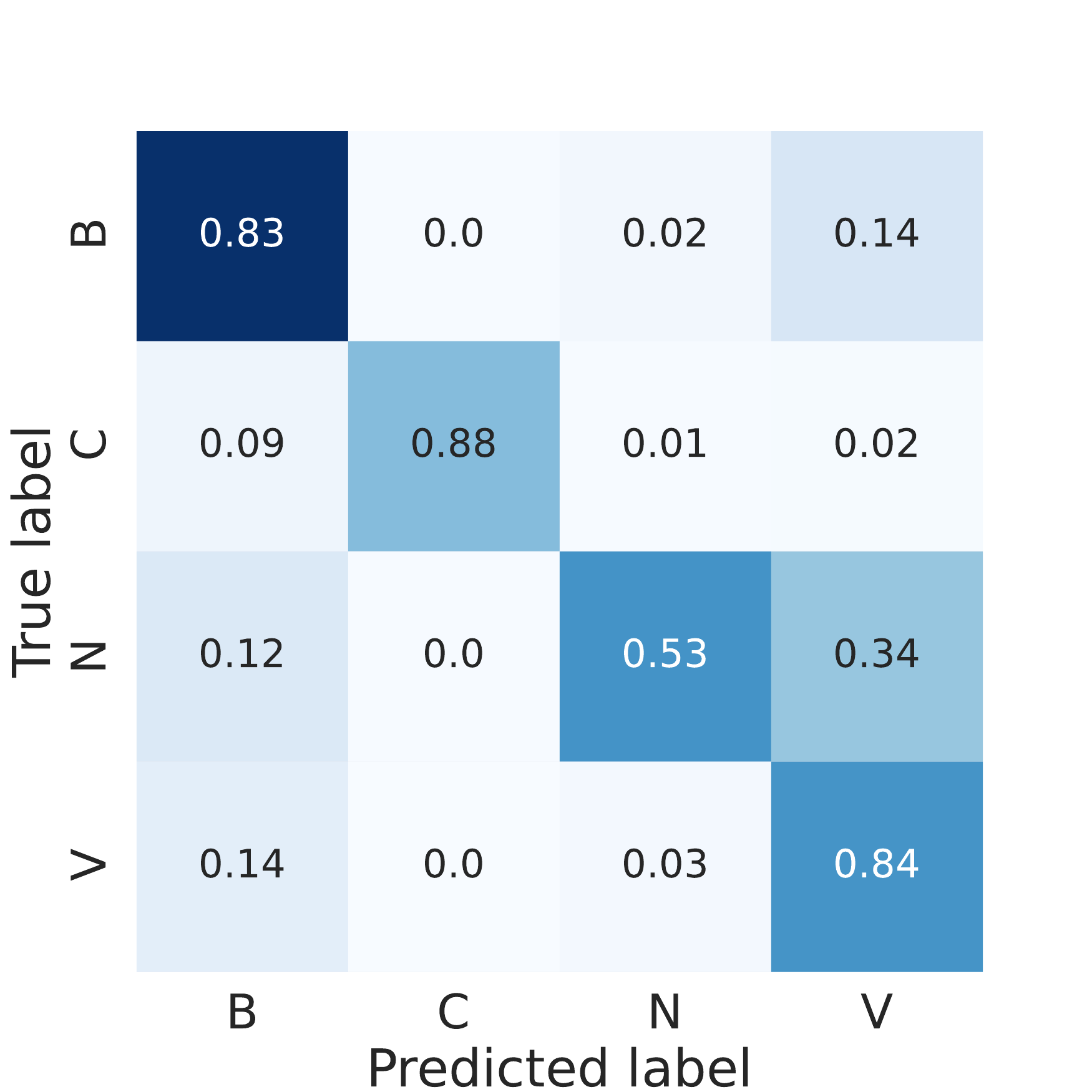}}
    \caption{Confusion matrices computed for AlexNet (a), DenseNet (b), GoogleNet (c), MnasNet (d), MobileNet v2 (e), MobileNet v3 (Large) (f), ResNet50 (g), ResNext (h), ShuffleNet (i), SqueezeNet (j), VGG16 (k) and Wide ResNet50 (l), computed on the test set. Labels N, B, V, C correspond to normal, bacteria, virus and COVID-19 classes, respectively.}
    \label{fig:confusion_matrices}
\end{figure*}

\begin{figure*}[t]
\captionsetup[subfigure]{labelformat=empty}
    \centering
    \subfloat{\includegraphics[width=.245\textwidth]{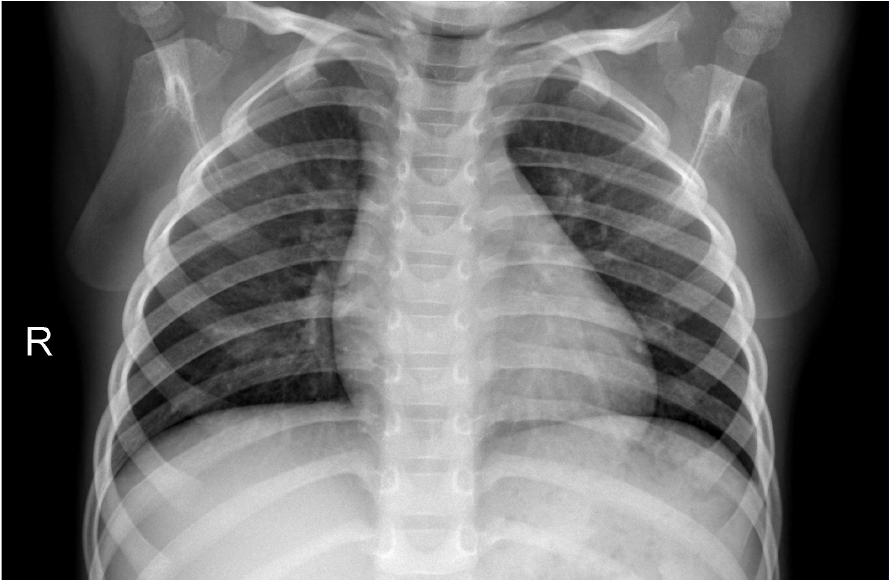}}
    \hfil
    \subfloat{\includegraphics[width=.245\textwidth]{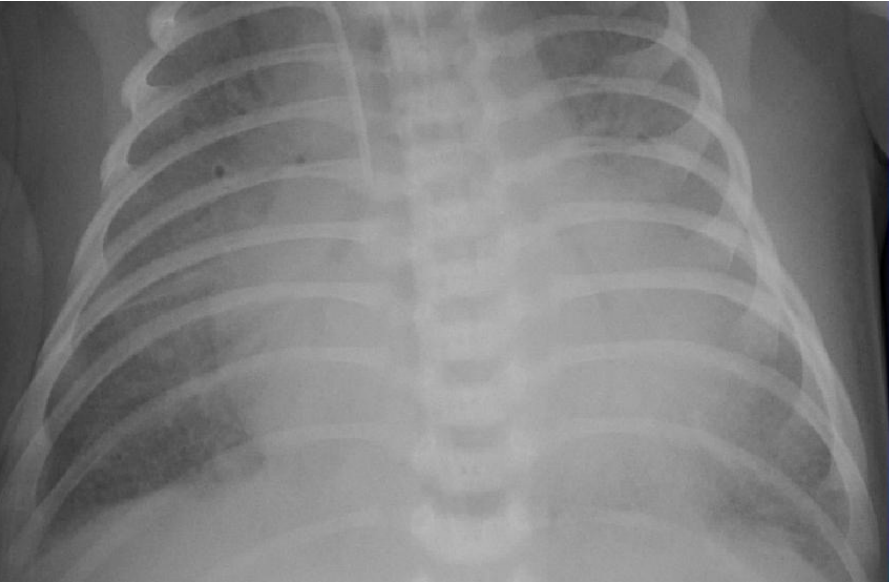}}
    \hfil
    \subfloat{\includegraphics[width=.245\textwidth]{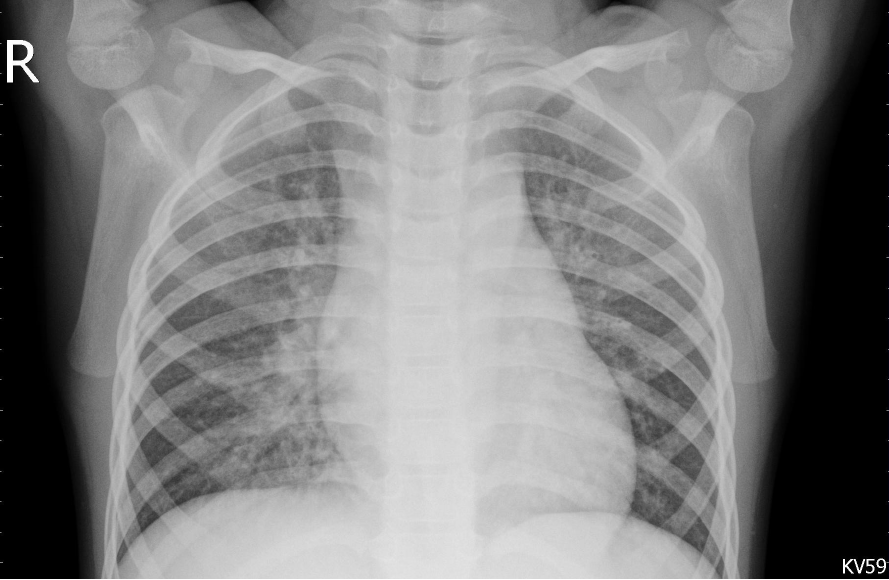}}
    \hfil
    \subfloat{\includegraphics[width=.245\textwidth]{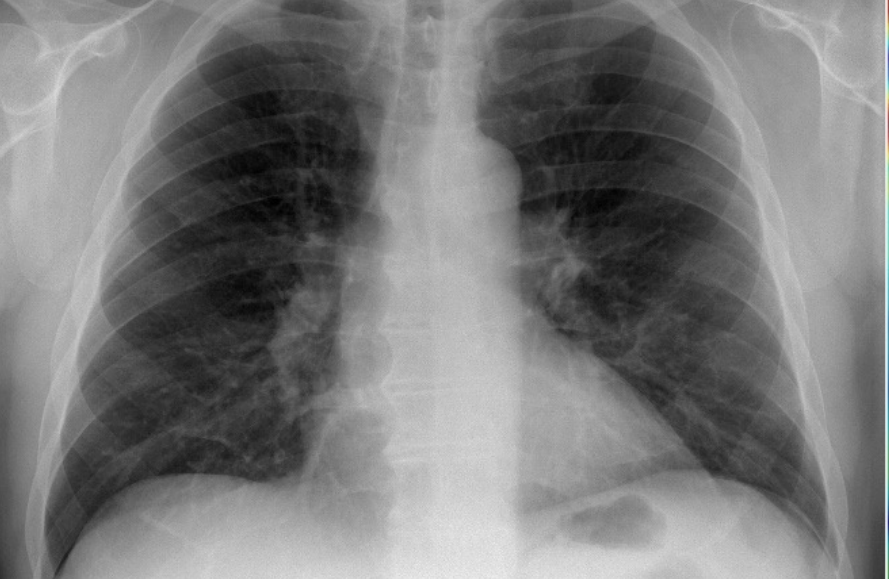}}
    \\[-2ex]
    
    \subfloat{\includegraphics[width=.245\textwidth]{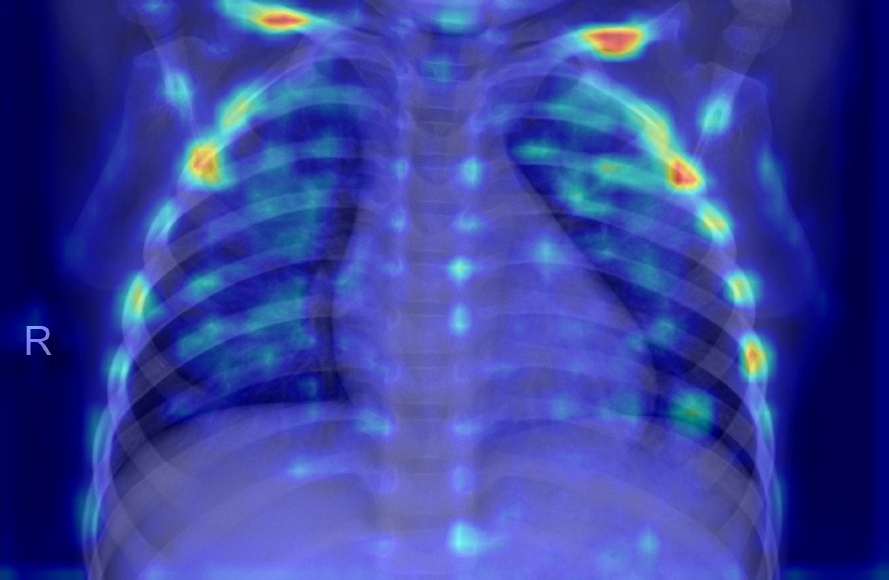}}
    \hfil
    \subfloat{\includegraphics[width=.245\textwidth]{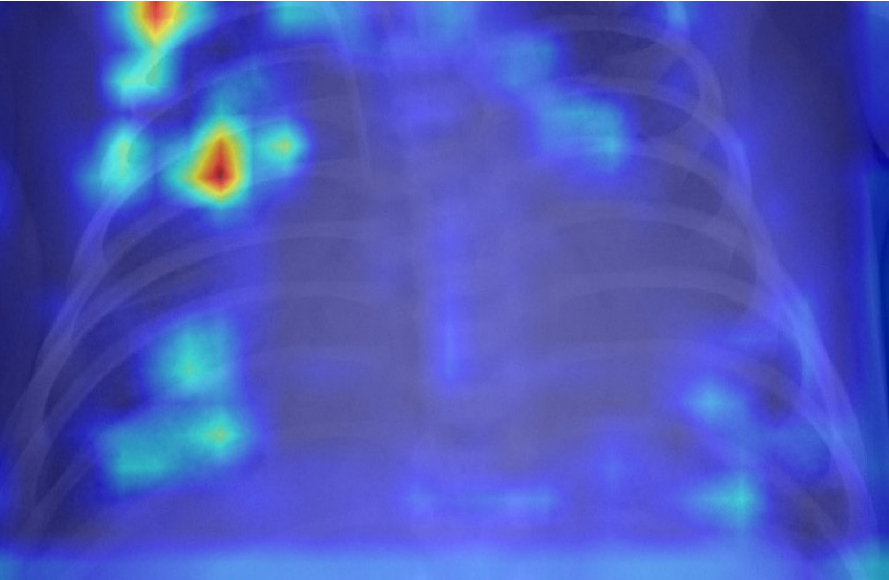}}
    \hfil
    \subfloat{\includegraphics[width=.245\textwidth]{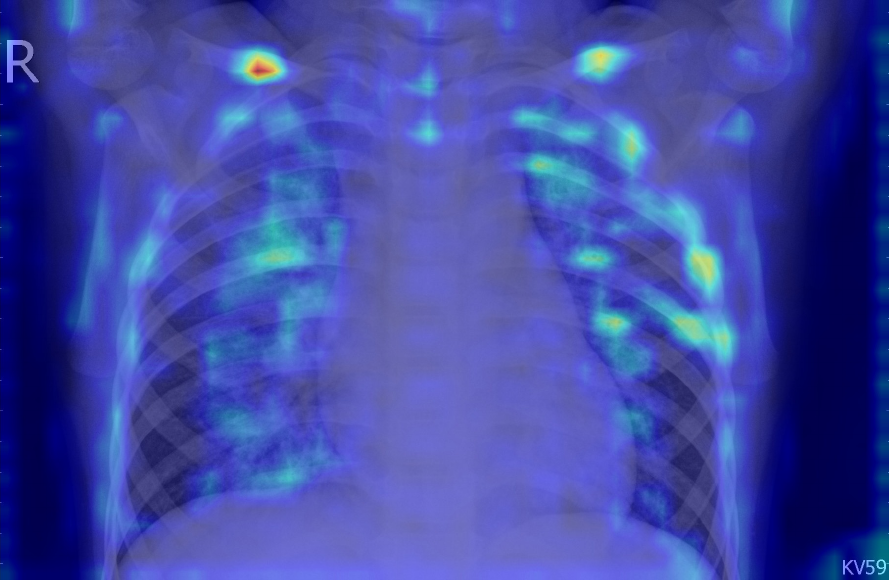}}
    \hfil
    \subfloat{\includegraphics[width=.245\textwidth]{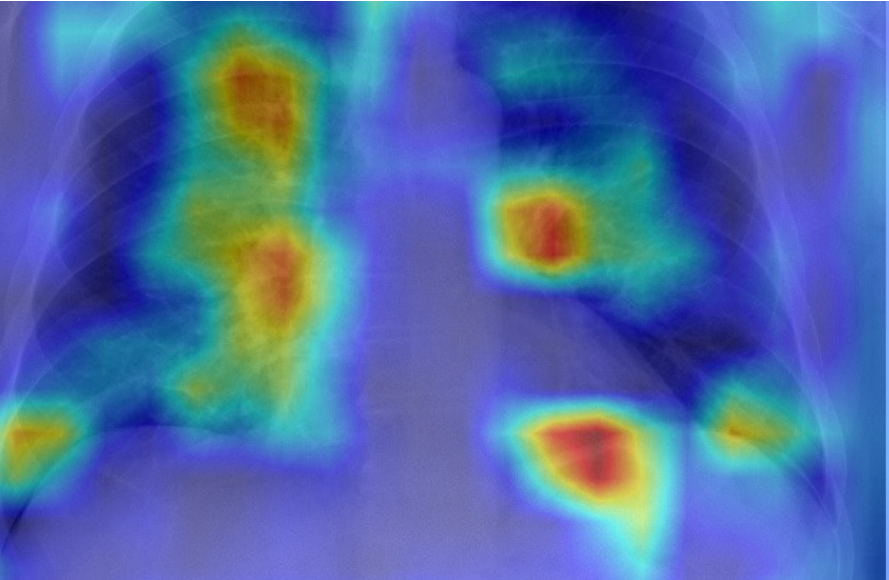}}
    \\[-2ex]
    
    \subfloat[(a)]{\includegraphics[width=.245\textwidth]{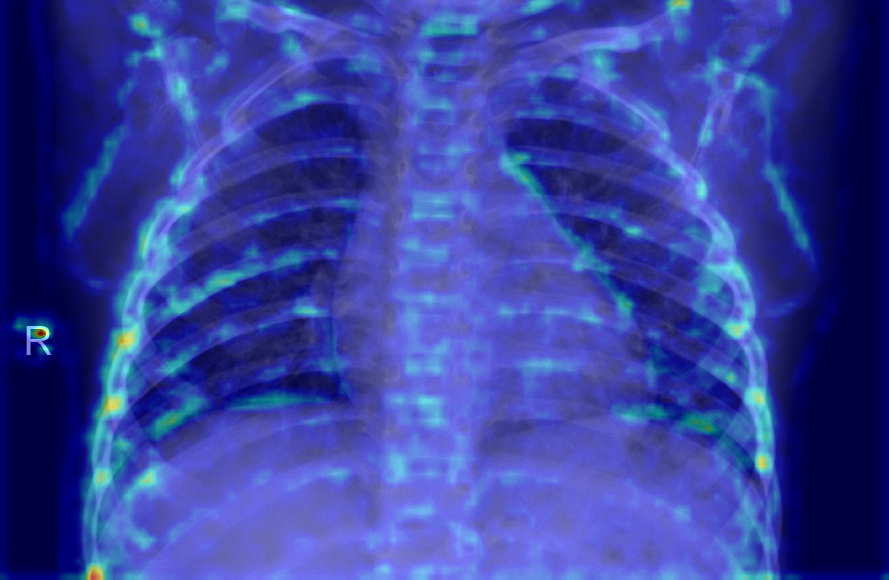}}
    \hfil
    \subfloat[(b)]{\includegraphics[width=.245\textwidth]{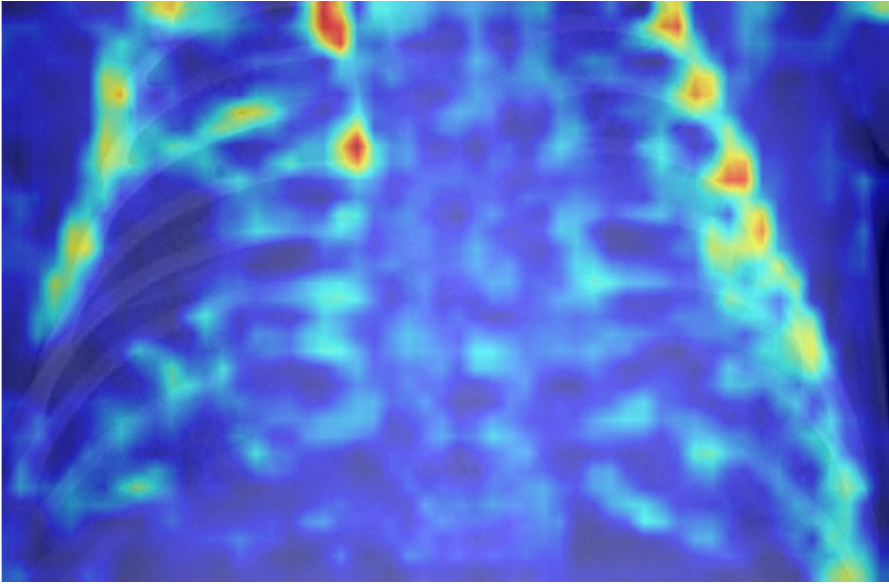}}
    \hfil
    \subfloat[(c)]{\includegraphics[width=.245\textwidth]{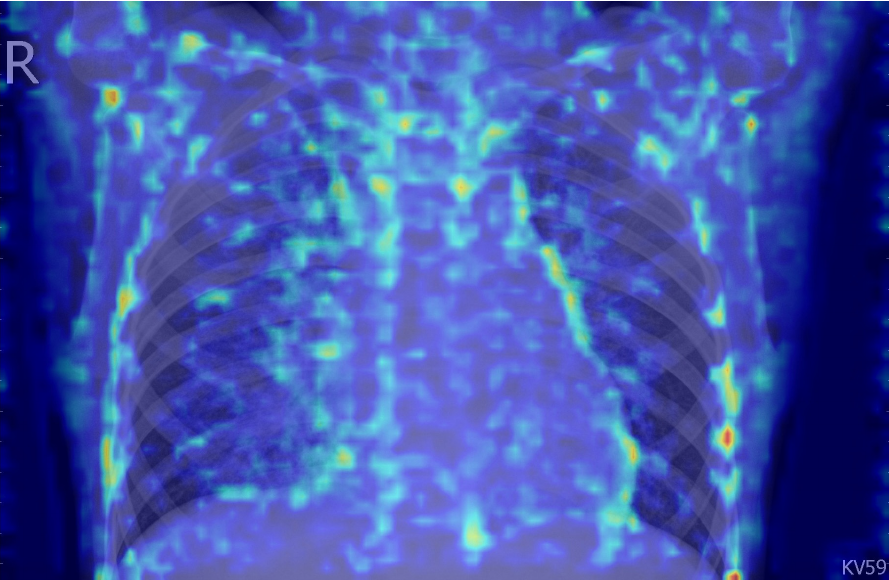}}
    \hfil
    \subfloat[(d)]{\includegraphics[width=.245\textwidth]{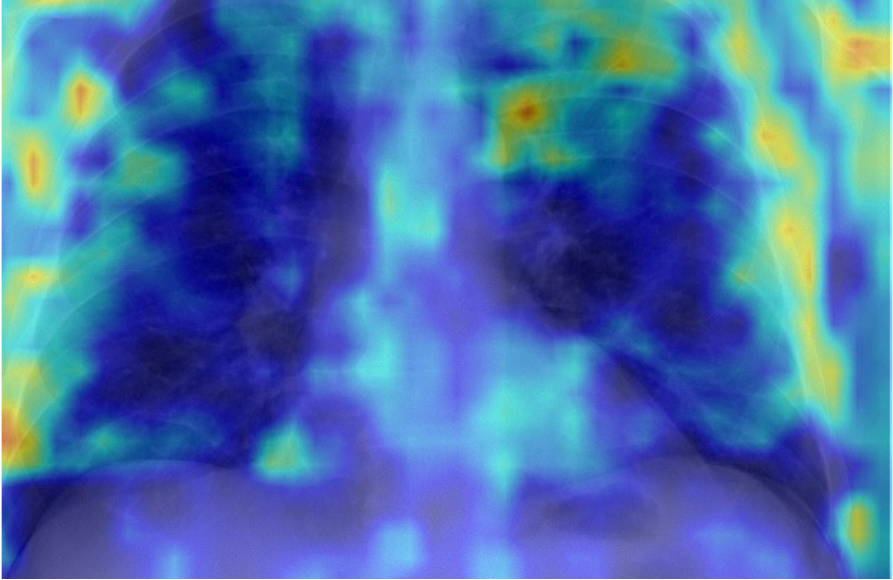}}
    \hfil
    \caption{Chest-x-ray inference samples showing a healthy patient (a), pneumonia from a bacteria infection (b), pneumonia from a virus infection (c), and pneumonia from a COVID-19 case (d). Input images are reported in the first row, while Grad-CAM overlaid images for MobileNet v3 and SqueezeNet are reported in the second and third row, respectively.}
    \label{fig:gradcam_output}
\end{figure*}

\begin{figure*}[t]
    \centering
    \subfloat[]{\includegraphics[width=.245\textwidth]{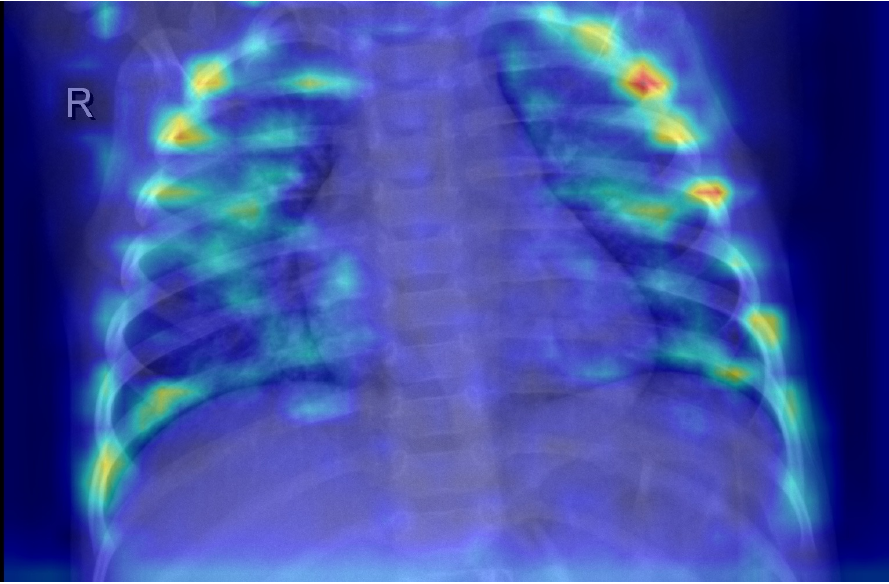}}
    \hfil
    \subfloat[]{\includegraphics[width=.245\textwidth]{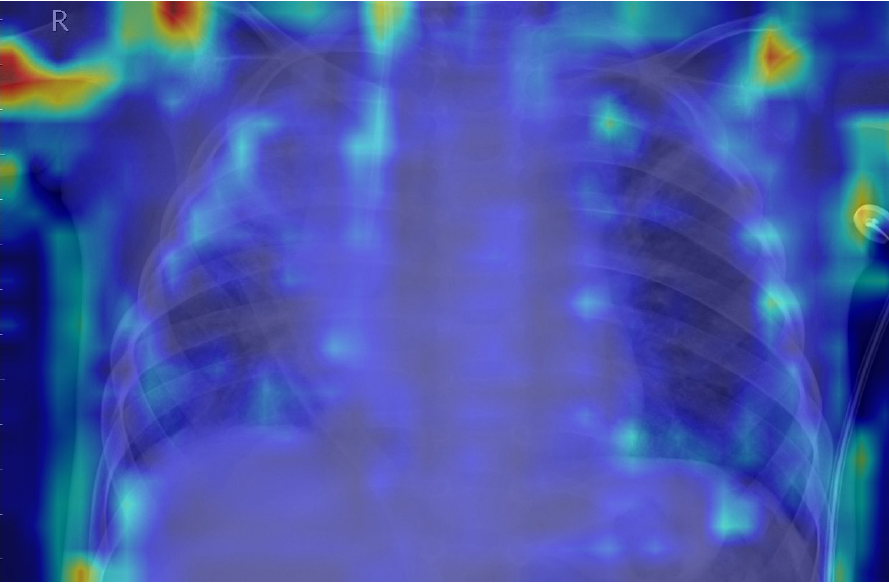}}
    \hfil
    \subfloat[]{\includegraphics[width=.245\textwidth]{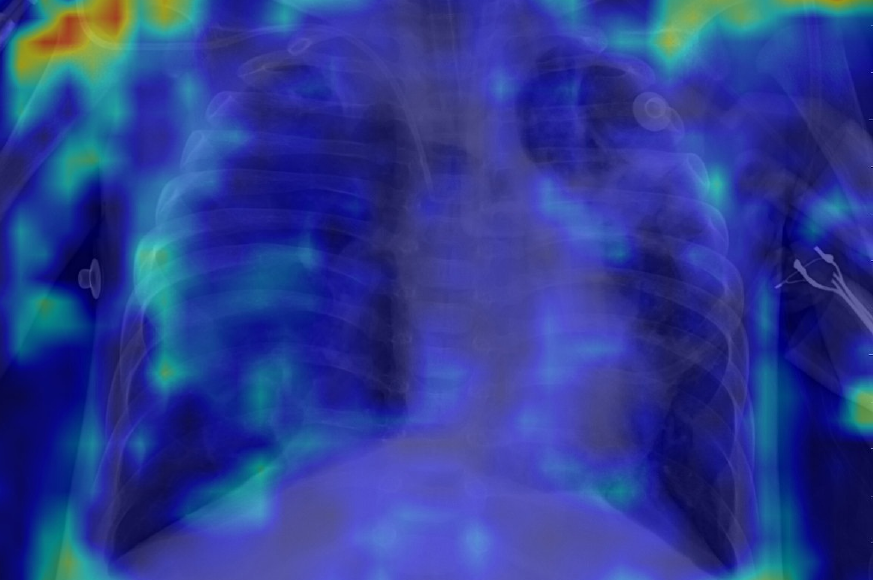}}
    \hfil
    \subfloat[]{\includegraphics[width=.245\textwidth]{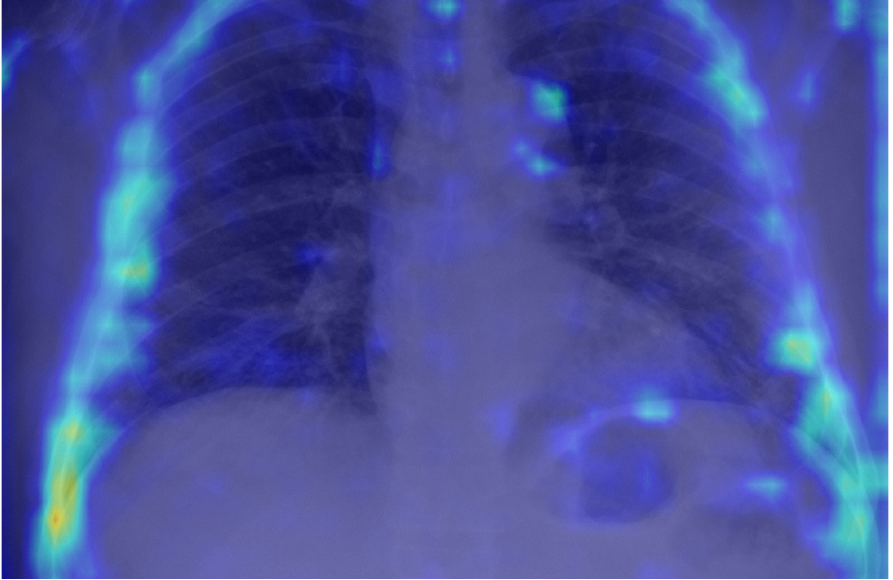}}
    \hfil
    \caption{Chest-x-ray misclassified samples showing a healthy patient (a), pneumonia from a bacteria infection (b), pneumonia from a virus infection (c), and pneumonia from a COVID-19 case (d), overlaid with Grad-CAM algorithm.}
    \label{fig:gradcam_misclassified}
\end{figure*}

\subsection{Performance Evaluation}\label{subsec:performance_evaluation}
In this section, we evaluate all of the presented models both quantitatively and qualitatively by means of sheer performances, for the former, and through an explainable visualization of the output, for the latter. 

Concerning the quantitative assessment, a graphical representation of the accuracy metric is illustrated in Fig.~\ref{fig:test_acc}, while relevant classification metrics are reported in Table \ref{tab:results}.
As shown, almost all networks can achieve satisfactory results with a score higher than 70\% across all metrics, indicating that transfer learning can be an important tool for the early screening of new diseases. Moreover, all models achieve high specificity scores, implying that, for each class, there generally is a low number of false positives, i.e., pneumonia sources tend not to be mixed during predictions.
More interestingly, the top performing architecture is the MobileNet v3, even though it is one of the models with the lowest number of parameters since it is optimized for mobile devices. This outcome highlights that more recent and advanced architectures can have an edge over older ones (e.g., AlexNet) and, as a consequence, may perform better using significantly fewer parameters without loss of generality.
While networks with less parameters can still obtain remarkable performances, the opposite result can also be observed. Examples of this behavior are the MnasNet and, in particular, SqueezeNet, which do not perform well during inference as they cannot effectively transfer previous knowledge to the new task due to both their architecture and parameters number; an expected outcome since these networks are mainly focused on significantly reducing pre-existing models sizes and can have a lower accuracy as a trade-off. Furthermore, the reduced performances are also confirmed by the average accuracies and losses, which are reported in Fig.~\ref{fig:acc_conv} and Fig.~\ref{fig:loss_conv}, respectively. As can be observed, most networks start converging well before the 40-th epoch, apart from the best performing DenseNet and MobileNet v3 that kept improving across the entire training procedure. This implies that previous knowledge, although useful, can saturate fast during its transferal. 
What is more, the SqueezeNet architecture actually diverges when adapting its classifier to the new task, thus resulting in loss values outside the reported range, and further justifying the low performance obtained on the pneumonia classification task. This outcome suggests that using an extremely low number of parameters can result in issues when exploiting the transfer learning paradigm, especially in the case of complex tasks such as the classification of chest-x-ray images. These results are also confirmed by the AUROCs reported in Fig.~\ref{fig:auroc}. As shown, most models obtain remarkable areas on each class, with the highest one being associated consistently with COVID-19, most likely due to the extreme effects of pneumonia spread inside the lungs derived from a SARS-CoV-2 infection. Such an outcome indicates that the models can correctly distinguish healthy patients from those affected by pneumonia and, for the latter, also individuate the correct source, further highlighting the effectiveness of the transfer learning paradigm to classify new illnesses. Concluding, notice that the only exception is the SqueezeNet architecture, i.e., Fig.~\ref{fig:auroc}.(j) since it did not converge during training.

On a different note, we performed several experiments to evaluate the robustness of each model by reducing the training set size. In particular, we used 50\%, 20\%, and 10\% of the collected training images to simulate a real case scenario where chest-x-rays would be limited upon the illness discovery. The results, obtained on the same test set described in Table~\ref{tab:hparams}, are reported in Table~\ref{tab:results-50}, Table~\ref{tab:results-20}, and Table~\ref{tab:results-10} for the training set reduced to 50\%, 20\%, and 10\%, respectively. Notice that this percentage is computed per class, following the training set described in Table~\ref{tab:hparams}, to ensure samples from each source would be included in the training set. As expected, performances naturally decrease for all networks by reducing the training set size, with the smallest size, i.e., 10\%, resulting in the lowest scores across all metrics. More interestingly, the best performing model with the entire training set, i.e., MobileNet v3, suffers from the highest decrease in performance, reaching almost 56\% less F1-score. On the one hand, this indicates a possible limit with the architecture itself, especially noticeable compared to its previous version MobileNet v2 which has a 17\% performance drop. On the other hand, it demonstrates that models with low performances can achieve higher results when provided with enough samples. Contrary to MobileNet v3, AlexNet has the smallest reduction in performance, with only a 3.39\% decrease in F1-score when using 10\% of the training set, suggesting that this convolutional network extracts robust features. In this context, another model that obtained remarkable scores across all metrics and training set sizes is the DenseNet architecture, implying that the internal configuration of a model can influence the computation of meaningful features, which is a relevant factor when analyzing newly discovered illnesses.

To conclude this quantitative evaluation and better appreciate the capabilities of knowledge transfer, confusion matrices for all models are shown in Fig.~\ref{fig:confusion_matrices}. As can be observed, excluding the diverged SqueezeNet architecture, i.e., Fig.~\ref{fig:confusion_matrices}.(j), predictions from the various models concentrate, as expected, on the corresponding matrix diagonal, indicating the right recognition of the four classes. On a different note, most of the errors are associated to the misclassification of the normal healthy class into the viral one. A behavior that can most likely be associated to artifacts present in the chest-x-ray, which might be confused for the diffused aspect of mild pneumonia cases from viral sources, in accordance with the findings of \cite{shah2017does}. Nevertheless, there is a robust COVID-19 recognition across the architectures, even though viruses, and especially other SARS pathogens, fall into a different category. This outcome highlights the effectiveness of transfer learning, and indicates that, for early diagnosis of new illnesses, it can be a practical strategy thanks to the pre-existing models prowess.

Regarding the qualitative evaluation, Grad-CAM overlaid images for the best and worst performing models, i.e., the MobileNet v3 and SqueezeNet architectures, are shown in Fig.~\ref{fig:gradcam_output}. As can be observed, the former model, reported on the second row, concentrates on the lungs for pneumonia cases deriving from bacteria, virus, and COVID-19, i.e., Fig.~\ref{fig:gradcam_output}.(b-d); while the architecture internal weights have a higher response, i.e., red areas on the Grad-CAM image, on both collar as well as rib cage bones for a normal chest-x-ray, i.e., Fig.~\ref{fig:gradcam_output}.(a). This outcome is most likely due to the pneumonia absence from the input, which resulted in the model focusing on other points of interest to learn the right category. Even more interesting, the activated areas in pneumonia images seem to confirm the illness propagation described in \cite{shah2017does,reynolds2010pneumonia}. Specifically, bacterial pathogens present a concentrated activation on specific spots, while viral sources present a diffused activation throughout the lungs. This effect is especially visible on COVID-19 pneumonia, i.e., Fig.~\ref{fig:gradcam_output}.(d), where bigger lungs areas have a higher response during inference, indicating that the model could also learn differences between viral and SARS-CoV-2 pathogens. Differently from the MobileNet v3, the SqueezeNet architecture tends to focus on bones and edges from all chest components by focusing on smaller details, most likely due to the smaller kernel filters of the network. Consequently, this model mixes normal chest-x-rays with either a bacterial or other generic viral pathogens, as expected from the diverged confusion matrix presented in Fig.\ref{fig:confusion_matrices}.(j). A similar issue for this second network, albeit to a lesser extent, is also presented when examining activated areas in the bacterial and COVID-19 images, i.e., Fig.~\ref{fig:gradcam_output}.(b) and Fig.~\ref{fig:gradcam_output}.(d). In both cases, the model responds to wider regions of the chest-x-ray in conjunction with rib cage bones, suggesting that the architecture is not analyzing pneumonia effects on lungs and resulting in the sensibly lower reported performances as well as training divergence. What is more, a comparable scenario can also be observed in Fig.~\ref{fig:gradcam_misclassified}, where misclassified images for the MobileNet v3, i.e., the best architecture on the presented dataset, are reported. As shown, for each category, the model has a strong response on other points of interest, such as collar or rib cage bones, as well as external lungs locations. Furthermore, for the healthy chest-x-ray, i.e., Fig.~\ref{fig:gradcam_misclassified}.(a), there are active areas within the lung in correspondence with image artifacts. Such behaviors indicate that, as mentioned, the architecture fine-tuning could be further improved, especially since it did not reach convergence on the training dataset. Moreover, it highlights the complexity of the task  since even proficient networks have difficulties handling these information-rich images. Regardless of these issues, qualitative experiments corroborate the quantitative effectiveness of transfer learning to diagnose pneumonia sources.

\section{Conclusion}\label{sec:conclusion}
In this paper, we presented a benchmark with 12 renowned deep neural network architectures modified to exploit the transfer learning paradigm. The scope of this study is to describe a comprehensive evaluation of different models when addressing the classification of a specific illness, in order to show both limitations and effectiveness of many remarkable architectures when applied to a new task. In particular, a dataset was organized from several public collections of chest-x-rays for the pneumonia classification derived by bacteria, generic viral or SARS-CoV-2 pathogens. 
Moreover, quantitative and qualitative experiments were reported to assess all models, which managed to reach up to 84.46\% average f1-score at inference time. As shown by the results, one of the smallest architectures, i.e., MobileNet v3, achieves the best performances and, when applying the Grad-CAM algorithm, it exhibits similar patterns with respect to the corresponding pneumonia source, i.e., high internal weights responses in concentrated or diffused areas for bacterial or viral and SARS-CoV-2 sources, respectively; demonstrating that previous knowledge does indeed help to address a new task, and suggesting that transfer learning can become a fundamental tool when diagnosing future unknown illnesses. Furthermore, the robustness of each model was assessed by reducing the training set size to 50\%, 20\%, and 10\% of its original dimension. The results highlighted both limitations and effectiveness of the tested architectures when provided with a reduced amount of samples. For instance, MobileNet v3 had a considerable 56\% F1-score decrease while AlexNet showed a reduction of 3.39\% when using the smallest training set.

As future work, an even larger collection will be organized to account for more variegate pulmonary diseases by merging other relevant public collections. In addition, further inquiries will be made by also evaluating other models as well as pre-processing techniques that could improve a given architecture performances by mitigating misclassification through the reduction, for instance, of artifacts in the input. Finally, we will develop generic AI-driven tools that can assist clinicians when diagnosing either known or newly discovered illnesses. 

\section*{Acknowledgments}
This work was supported by the MIUR under grant “Departments of Excellence 2018–2022” of the Sapienza University Computer Science Department and the ERC Starting Grant no. 802554 (SPECGEO).

\bibliography{biblio}

\end{document}